\documentclass[12pt]{article}
\usepackage{amssymb,amsfonts}
\usepackage{epsf,epsfig}
\def \la{\lambda}

\begin{document}
\baselineskip 18pt

\title{Three-magnon problem for exactly rung-dimerized spin ladders: from general outlook to Bethe Ansatze}
\author{P.~N.~Bibikov\\ \it Sankt-Petersburg State University}

\maketitle

\vskip5mm

\begin{abstract}
Three-magnon problem for exactly rung-dimerized spin ladder is
brought up separately at all total spin sectors. At first a
special duality transformation of the $\rm Schr\ddot odinger$
equation is found within general outlook. Then the problem is
treated within Coordinate Bethe Ansatze. A straightforward
approach is developed to obtain pure scattering states. At values
$S=0$ and $S=3$ of total spin the $\rm Schr\ddot odinger$ equation
has the form inherent in the $XXZ$ chain. For $S=1,2$ solvability
holds only in five previously found {\it completely integrable}
cases. Nevertheless a partial $S=1$ Bethe solution always exists
even for general non integrable model. Pure scattering states for
all total spin sectors are presented explicitly.
\end{abstract}

\section{Introduction}

Among other gapped 1D systems spin ladders were intensively
studied during the last 15 years experimentally, numerically and
theoretically (see Refs. in \cite{1}-\cite{3}). The interest is
accounted by their possibly relation to high temperature
superconductivity, variety of static and dynamical properties and
even an existence of several reliable compounds.

In the pioneering paper \cite{4} a spin ladder was suggested as a
double spin chain with Heisenberg interactions both across and
along the chains direction namely rung and leg exchanges related
to the couplings $J_r$ and $J_l$. It was also pointed that the
case,
\begin{equation}
J_r\gg J_l,
\end{equation}
has a principle interest because it belongs to the so called {\it
rung-dimerized phase} in which almost all spins are coupled into
rung-singlets (rung-dimers). In the purely Heisenberg reference
model this phase becomes exact only for $J_l=0$. However it is
always assumed that under the condition (1) the physical picture
does not change in common.

Soon it became clear that spin ladder Hamiltonian also admit a
term related to diagonal Heisenberg coupling as well as four spin
terms \cite{5}. At a first sight these new interactions seemed to
be complications for a theoretical analysis. However even in
\cite{6} it was noted that a special linear condition (the Eq.
(22) of the present paper) on the former and new coupling
constants guarantees (for rather big $J_r$) {\it exactness} of the
rung-dimerized ground state. Besides in this case all one- and
two-magnon states also may be obtained in explicit form
\cite{6},\cite{7}.

Unfortunately the rung-dimerization condition (22) has no reliable
atomic level interpretation, so there is no physical reason to
postulate it. Nevertheless it seems reasonable to suppose that for
strong rung exchange any deviations from the exact rung-dimerized
picture should be small and may be evaluated perturbatively. (In
more detail this question will be studied in a forthcoming paper.)
Under this point of view exactly rung-dimerized spin ladders are
the best reference models for treating the whole rung-dimerized
phase.

Some static and dynamic zero-temperature properties of exactly
rung-dimerized spin ladders were studied in a series of papers
\cite{7}-\cite{10}. Due to the gap it succeeded to describe Raman
scattering \cite{7}, magnetic phase transition \cite{8} and (for
asymmetric ladders) magnon decay \cite{9},\cite{10} utilizing only
one- and two-magnon spectrums. Three-magnon problem is less actual
for the $T=0$ physics (see however the papers \cite{11},\cite{12}
devoted to the $S=1$ Haldane chain and $O(3)$ nonlinear
$\sigma$-model).

Advancement into the $T>0$ region needs a knowledge of the whole
spectrum \cite{3},\cite{16}. However such level of clearness may
be achieved only for a rather limited list of the so called
integrable models \cite{3},\cite{13}-\cite{17}. The latter besides
are significant in heat transport phenomena \cite{18}.

But how to find an integrable model? How it may be identified from
a overwhelming majority of nonintegrable ones? The most direct way
is to express a treating Hamiltonian density as a derivative of
the corresponding $R$-matrix which satisfy the Yang-Baxter
equation. Solvability of this problem is governed by the
Reshetikhin condition \cite{17},\cite{19},\cite{20}. If the latter
is satisfied for a given local Hamiltonian density then the
corresponding $R$-matrix rather exist and may be obtained by an
analysis of power series \cite{20},\cite{21},\cite{22} or by some
Yang-Baxterization ansatze \cite{23}. In the present paper we
suggest an alternative approach based on solvability of the three
magnon problem in a framework of the Coordinate Bethe Ansatze
(CBA) \cite{24}.

The essence of the CBA method \cite{13},\cite{14} is an assumption
that any many-particle wave functions is in fact a linear
combination of terms produced by multiplications of one-particle
exponents. Namely, for a rung-dimerized spin ladder the one-magnon
wave function $\psi(n)={\rm e}^{ikn}$ \cite{6} is parameterized by
a real $0\leq k<2\pi$ (the wave number) and depends on an integer
$n$ (position of the triplet rung). A two magnon wave function
$\psi(m,n)$ ($m<n$) is linear combination of two exponents ${\rm
e}^{i(k_1m+k_2n)}$ and ${\rm e}^{i(k_2m+k_1n)}$ \cite{7} and so
depends on a pair of non equal parameters $k_1$ and $k_2$. For a
scattering state they both are real and one may put
\begin{equation}
0\leq k_1<k_2<2\pi,
\end{equation}
while for a bound state they are complex conjugate
\begin{equation}
k_2=\bar k_1.
\end{equation}

In this light it is seems reasonable to search for representation
of multi-magnon wave functions as sums of the Bethe exponents.
However even a subsequent development of this approach to the
three-magnon sector dashes on the problem of non integrability.

In order to reveal an origin of this obstacle let us at first turn
back to a two-magnon state. Total quasimomentum (wave number) and
energy of the latter are the sums
\begin{equation}
k=k_1+k_2,\quad E(k_1,k_2)=E_{magn}(k_1)+E_{magn}(k_2),
\end{equation}
where $E_{magn}(k)$ a single magnon energy. It is significant that
under the conditions (2) or (3) the mapping
\begin{equation}
k_1,k_2\longrightarrow k,E,
\end{equation}
given by (4) is uniqually (up to an exchange $k_1\leftrightarrow
k_2$) reversible. However for three magnons the situation is
drastically different. Indeed a system of relations
\begin{equation}
k=k_1+k_2+k_3,\quad E=E_{magn}(k_1)+E_{magn}(k_2)+E_{magn}(k_3),
\end{equation}
can define an infinite number of triples ($k_1,k_2,k_3$). As a
result a three-magnon wave function related to the pair $(k,E)$
should contain in general an {\it infinite} number of exponential
terms related to different solutions of the system (6). Evidently
such three magnon problem is practically unsolvable.

The above obstacle may be overcame by existence a first integral
(a translationary invariant operator commuting with the
Hamiltonian) which produces the third condition additional to (6).
An integrable system has an infinite number of such commuting in
pairs first integrals and may be solved in all multi-particle
sectors. It is significant that within the CBA a difference
between integrability and non-integrability manifests just at the
three particle level. As a consequence of this fact one may
consider solvability the of three-particle problem as an
alternative integrability test.

In the present paper we study three magnon sector of a
rung-dimerized symmetric spin ladder. At first we briefly analyze
the problem in general outlook and only afterwards turn to CBA.
Motivation of such approach is the following argumentation.
Usually CBA is treated as a successful ad hoc conjecture which
allows to obtain in a rather straightforward manner all
multi-particle states for a given an quantum integrable model.
However the reference one is not integrable at general values of
coupling constants. As a result (it will be shown below in detail)
the CBA approach is applicable only in five special integrable
cases.

The calculations are performed separately in the sectors $S=0,1,2$
(the $S=3$ sector is similar to the $S=0$ one) of total spin. At
$S=0$ ($S=3$) the system of equations on Bethe amplitudes has a
well known form inherent in the $XXZ$ spin chain and so is
completely solvable for all values of coupling constants. For
$S=1$ and $S=2$ a complete solvability takes place only in the
five integrable cases obtained earlier \cite{21} within the
Yang-Baxter framework. However even in the general nonintegrable
case there is a special (very complicated) solution in the $S=1$
sector. Its interpretation remains unclear.

The plan of the paper is the following. In Sect. 2 we represent
the spin ladder Hamiltonian in the most tractable form for which
the rung-dimerized condition is evident. In Sect. 3 we show that
the Bethe form of the two-magnon wave function readily follows
from a straightforward treatment of the $\rm Shr\ddot odinger$
equation. In Sect.4 treating within general framework the $S=0$
($S=3$) sector we reveal a duality transformation of wave function
(generalized in Sect. 5 and 6 for $S=1,2$) and show that the Bethe
Ansatze readily follows from the factorized (Fourier)
substitution. We also obtain a classification (generalized in
Sect. 5 and 6 for $S=1,2$) of Bethe three-magnon states related to
complex wave numbers. Pure scattering states obtained within a
straightforward approach developed in Sect. 4,5,6 are presented in
the Appendix. In Sect. 7 we show that the revealed CBA solvability
is in one to one correspondence with the integrability revealed
earlier within the Yang-Baxter framework \cite{21}. We also
present the corresponding R-matrices. In Sect. 8 within CBA we
describe action of the $S_3$ permutation group in all total spin
sectors. This symmetry as well as duality described in Sect. 5 and
6 is used in the Appendix for more compact representation of Bethe
states.

Since the ground state of the model has a simple factorized form
we treat it only in the infinite volume limit. Analogous approach
to the ferromagnetic $XXZ$ chain was developed in \cite{14}.

\section{The spin ladder Hamiltonian}

Before presenting the spin ladder Hamiltonian let us introduce the
following local operators
\begin{eqnarray}
{\bf\Psi}_n&=&\frac{1}{2}({\bf S}_{1,n}-{\bf S}_{2,n})-i{[}{\bf
S}_{1,n}\times{\bf
S}_{2,n}{]},\nonumber\\
{\bf\tilde\Psi}_n&=&\frac{1}{2}({\bf S}_{1,n}-{\bf
S}_{2,n})+i{[}{\bf S}_{1,n}\times{\bf S}_{2,n}{]},
\end{eqnarray}
(we use the notation ${\bf\tilde\Psi}_n$ instead of more
convenient ${\bf\Psi}^*_n$ or ${\bf\Psi}^{\dagger}_n$ only in
order to avoid such rather cumbersome notations as
$({\bf\Psi}^a_n)^*$). Here ${\bf S}_{1,n}$ and ${\bf S}_{2,n}$ are
local spin operators associated with $n$-th rung. They may be
expressed from ${\bf\Psi}_n$ and ${\bf\tilde\Psi}_n$ as follows
\begin{eqnarray}
{\mathbf
S}_{1,n}&=&\frac{1}{2}\Big({\bf\Psi}_n+{\bf\tilde\Psi}_n-i[{\bf\tilde\Psi}_n\times{\bf\Psi_n}]\Big),\nonumber\\
{\mathbf
S}_{2,n}&=&\frac{1}{2}\Big(-{\bf\Psi}_n-{\bf\tilde\Psi}_n-i[{\bf\tilde\Psi}_n\times{\bf\Psi_n}]\Big).
\end{eqnarray}
The representation (8) is similar to the one suggested in
\cite{25} but in fact is not identical to it. Really the analogs
of ${\bf\Psi}_n$ and ${\bf\tilde\Psi}_n$ treated in \cite{25} act
in an extended vector space. That is why for example the "inverse"
representation (7) fails for them.

It may be readily proved that
\begin{equation}
[{\bf\Psi}_n,Q_n]={\bf\Psi}_n,\quad
[{\bf\tilde\Psi}_n,Q_n]=-{\bf\tilde\Psi}_n,
\end{equation}
where
\begin{equation}
Q_n=\frac{1}{2}{\mathbf S}_n^2,\quad {\mathbf S}_n={\mathbf
S}_{1,n}+{\mathbf S}_{2,n},
\end{equation}

Let $|0\rangle_n$ and $|1\rangle_n$ be correspondingly singlet and
triplet states associated with $n$-th rung. From (10) follows that
\begin{equation}
Q_n|0\rangle_n=0,\qquad Q_n|1\rangle_n=|1\rangle_n,
\end{equation}
so the local operator $Q_n$ is projector on the $n$-th rung
triplet sector. Then according to Eq. (9) the two triples
${\bf\tilde\Psi}_n$ and ${\bf\Psi}_n$ may be treated as
rung-triplet creation-annihilation operators. Namely the tripe
$|1\rangle^a_n$ ($a=x,y,z$) for which
\begin{equation}
{\bf\tilde\Psi}^a_n|0\rangle_n=|1\rangle^a_n,\quad{\bf\tilde\Psi}^a_n|1\rangle^b_n=0,\quad
{\bf\Psi}^a_n|0\rangle_n=0,\quad
{\bf\Psi}^a_n|1\rangle^b_n=\delta_{ab}|0\rangle_n,
\end{equation}
gives the following representation of the total rung-spin
\begin{equation}
{\mathbf S}_n^a|1\rangle_n^b=i\epsilon_{abc}|1\rangle_n^c,
\end{equation}
($\epsilon_{abc}$ is the Levi-Chivita tensor). Parallel with (12)
we shall use the triple
\begin{equation}
|1\rangle^j_n={\bf\tilde\Psi}^j_n|0\rangle_n,\quad{\mathbf
S}_n^z|1\rangle^j_n=j|1\rangle^j_n,\quad j=-1,0,1.
\end{equation}
related to operators
\begin{equation}
{\bf\tilde\Psi}^{\pm1}_n\equiv\frac{1}{\sqrt{2}}\Big({\bf\tilde\Psi}^x_n\pm
i{\bf\tilde\Psi}^y_n\Big),\quad
{\bf\tilde\Psi}^0_n\equiv{\bf\tilde\Psi}^z_n.
\end{equation}

It seems reasonable to represent the Hamiltonian density
$H_{n,n+1}$ for general spin ladder Hamiltonian
\begin{equation}
{\hat H}=\sum_nH_{n,n+1},
\end{equation}
in the following form
\begin{eqnarray}
H_{n,n+1}&=&J_1(Q_n+Q_{n+1})+
J_2({\bf\Psi}_n\cdot{\bf\tilde\Psi}_{n+1}+{\bf\tilde\Psi}_n\cdot{\bf\Psi}_{n+1})\nonumber\\
&+&J_3Q_nQ_{n+1}+J_4{\bf S}_n\cdot{\bf S}_{n+1}+J_5({\bf S}_n\cdot{\bf S}_{n+1})^2\nonumber\\
&+&J_6({\bf\tilde\Psi}_n\cdot{\bf\tilde\Psi}_{n+1}+{\bf\Psi}_n\cdot{\bf\Psi}_{n+1}).
\end{eqnarray}
Up to a constant this representation is equivalent to the standard
one \cite{1}-\cite{6}
\begin{equation}
H_{n,n+1}=J_rH^r_{n,n+1}+J_lH^l_{n,n+1}
+J_dH^d_{n,n+1}+J_{rr}H^{rr}_{n,n+1}
+J_{ll}H^{ll}_{n,n+1}+J_{dd}H^{dd}_{n,n+1},
\end{equation}
where
\begin{eqnarray}
H^r_{n,n+1}&=&\frac{1}{2}({\bf S}_{1,n}\cdot{\bf S}_{2,n}+{\bf
S}_{1,n+1}\cdot{\bf S}_{2,n+1}),\quad H^l_{n,n+1}={\bf
S}_{1,n}\cdot{\bf S}_{1,n+1}+{\bf S}_{2,n}\cdot{\bf
S}_{2,n+1},\nonumber\\
H^d_{n,n+1}&=&{\bf S}_{1,n}\cdot{\bf S}_{2,n+1}+{\bf
S}_{2,n}\cdot{\bf S}_{1,n+1},\quad H^{rr}_{n,n+1}=({\bf
S}_{1,n}\cdot{\bf S}_{2,n})({\bf S}_{1,n+1}\cdot{\bf
S}_{2,n+1}),\nonumber\\
H^{ll}_{n,n+1}&=&({\bf S}_{1,n}\cdot{\bf S}_{1,n+1})({\bf
S}_{2,n}\cdot{\bf S}_{2,n+1}),\quad H^{dd}_{n,n+1}=({\bf
S}_{1,n}\cdot{\bf S}_{2,n+1})({\bf S}_{2,n}\cdot{\bf S}_{1,n+1}),
\end{eqnarray}
and
\begin{eqnarray}
J_1&=&\frac{1}{4}\Big(2J_r-3J_{rr}-J_{ll}-J_{dd}\Big),\nonumber\\
J_2&=&\frac{1}{8}\Big(4(J_l-J_d)+J_{ll}-J_{dd}\Big),\nonumber\\
J_3&=&J_{rr},\nonumber\\
J_4&=&\frac{1}{8}\Big(4(J_l+J_d)+J_{ll}+J_{dd}\Big),\nonumber\\
J_5&=&\frac{1}{4}\Big(J_{ll}+J_{dd}\Big),\nonumber\\
J_6&=&\frac{1}{8}\Big(4(J_l-J_d)-J_{ll}+J_{dd}\Big).
\end{eqnarray}

It was suggested in \cite{5} that only the case
\begin{equation}
J_{rr}=J_{ll}=-J_{dd},
\end{equation}
(or equivalently $J_5=0$, $J_6=J_2-J_3/2$) has a reliable
interest. However since spin ladders with failed condition (21)
also are currently studied \cite{3} and we shall not require it.

From (9) and (17) directly follows that for
\begin{equation}
J_6=0\Leftrightarrow J_{ll}-J_{dd}=4(J_l-J_d),
\end{equation}
(triplet-rungs pair creation-annihilation processes are
suppressed) there holds
\begin{equation}
[\hat H,\hat Q]=0.
\end{equation}
Here the global operator
\begin{equation}
\hat Q=\sum_nQ_n,
\end{equation}
according to (11) may be treated as a number operator for
rung-triplets. The commutation relation (23) results in splitting
of the Hilbert space on an infinite sum of eigenspaces related to
different eigenvalues of $\hat Q$. In particularly for rather
strong $J_1$ the (zero energy) ground state of the model has a
simple tensor-product form \cite{6}
\begin{equation}
|0\rangle=\prod_n\otimes|0\rangle_n.
\end{equation}
At the same time the physical Hilbert space is subdivided into a
direct sum of magnon sectors
\begin{equation}
{\cal H}=\sum_{m=0}^{\infty}{\cal H}^m,\quad \hat Q|_{{\cal
H}^m}=m.
\end{equation}
Only this special case (Eq. (22) and rather strong $J_1$) will be
studied in the present paper. Additionally we shall imply that
$J_2\neq0$. The completely diagonal frustrated model related to
$J_2=0$ or equivalently $J_d=J_l$ (in this case the Hamiltonian
density (17) may be expressed only in terms of $Q_n$ and ${\bf
S}_n$) was studied in details in \cite{26}. Besides one may assume
that
\begin{equation}
J_2>0\Leftrightarrow J_l>J_d.
\end{equation}
Indeed the case $J_2<0$ can be reduced to (27) by use of the
following exchange of the coupling constants
\begin{equation}
J_l\leftrightarrow J_d,\qquad J_{ll}\leftrightarrow J_{dd},
\end{equation}
related to permutation of spins on all even (odd) rungs.

\section{One- and two-magnon states}

Taking into account (17), (11), (12) and (14) one gets the local
formulas
\begin{eqnarray}
H_{n,n+1}\dots|1\rangle_n|0\rangle_{n+1}\dots&=&
J_1\dots|1\rangle_n|0\rangle_{n+1}\dots+
J_2\dots|0\rangle_n|1\rangle_{n+1}\dots,\nonumber\\
H_{n-1,n}\dots|0\rangle_{n-1}|1\rangle_n\dots&=&
J_1\dots|0\rangle_{n-1}|1\rangle_n\dots+
J_2\dots|1\rangle_{n-1}|0\rangle_n\dots,
\end{eqnarray}
and
\begin{eqnarray}
H_{n,n+1}\dots|1\rangle_n^a|1\rangle_{n+1}^a\dots&=&
\varepsilon_0\dots|1\rangle_n^a|1\rangle_{n+1}^a\dots,\nonumber\\
H_{n,n+1}\epsilon_{abc}\dots|1\rangle_n^b|1\rangle_{n+1}^c\dots&=&
\varepsilon_1\epsilon_{abc}\dots|1\rangle_n^b|1\rangle_{n+1}^c\dots,\quad (a,b,c=x,y,z)\nonumber\\
H_{n,n+1}\dots|1\rangle_n^+|1\rangle_{n+1}^+&\dots=&
\varepsilon_2\dots|1\rangle_n^+|1\rangle_{n+1}^+\dots
\end{eqnarray}
Here
\begin{equation}
\varepsilon_S\equiv2(J_1+J_2\Delta_S),
\end{equation}
and
\begin{eqnarray}
\Delta_0&=&\frac{J_3-2J_4+4J_5}{2J_2}=\frac{4(J_d-2J_l)+2J_{rr}+3J_{ll}}{4(J_l-J_d)},\nonumber\\
\Delta_1&=&\frac{J_3-J_4+J_5}{2J_2}=\frac{4(J_{rr}-J_l)+J_{ll}}{8(J_l-J_d)},\nonumber\\
\Delta_2&=&\frac{J_3+J_4+J_5}{2J_2}=\frac{4(2J_d-J_l)+4J_{rr}+3J_{ll}}{8(J_l-J_d)}.
\end{eqnarray}

From (30) may be readily obtained a useful formula
\begin{eqnarray}
H_{n,n+1}\dots|1\rangle_n^a|1\rangle_{n+1}^b\dots&=&\Big(2J_1+J_3+J_5\Big)\dots|1\rangle_n^a|1\rangle_{n+1}^b\dots
+J_4\dots|1\rangle_n^b|1\rangle_{n+1}^a\dots\nonumber\\
&+&
\delta_{ab}\Big(J_5-J_4\Big)\dots|1\rangle_n^c|1\rangle_{n+1}^c\dots,\quad(a,b,c=x,y,z).
\end{eqnarray}

Turning to excitation states we notice that an explicit form of a
one-magnon state
\begin{equation}
|1,k\rangle=\sum_ne^{ikn}\Big(\prod_{m=-\infty}^{n-1}\otimes|0\rangle_m\Big)\otimes
|1\rangle_n\otimes\Big(\prod_{m=n+1}^{\infty}\otimes|0\rangle_m\Big).
\end{equation}
directly follows from (23) and translation symmetry
\begin{equation}
\hat P|1,k\rangle={\rm e}^{-ik}|1,k\rangle.
\end{equation}
Here $\hat P$ is the translation operator
\begin{equation}
\hat
P\prod\otimes|\chi(n)\rangle_n=\prod\otimes|\chi(n)\rangle_{n+1},\qquad\chi(n)=0,1.
\end{equation}
The corresponding dispersion
\begin{equation}
E_{magn}(k)=2(J_1+J_2\cos k),
\end{equation}
readily follows from (29).

Since the $\hat Q=2$ sector is subdivided on the total spin
$S=0,1,2$ subsectors we denote at once a two-magnon state with
total spin $S$ and wave vector $k$ as $|2,S,k\rangle$. The
following general representations for the two-magnon states
\begin{eqnarray}
|2,0,k\rangle&=&\sum_{m<n} {\rm e}^{ik(m+n)/2}a_0(k,n-m)
\dots|1\rangle^a_m\dots|1\rangle^a_n\dots,\nonumber\\
|2,1,k\rangle^a&=&\varepsilon_{abc}\sum_{m<n} {\rm
e}^{ik(m+n)/2}a_1(k,n-m)
\dots|1\rangle^b_m\dots|1\rangle^c_n\dots,\nonumber\\
|2,2,k\rangle^{+2}&=&\sum_{m<n} {\rm e}^{ik(m+n)/2}a_2(k,n-m)
\dots|1\rangle^+_m\dots|1\rangle^+_n\dots,
\end{eqnarray}
agree with the rotational and translation (35) symmetries. From
Eq. (38) by "$\dots$" we denote an appropriate tensor product of
{\it rung-singlets} (similar to products in (34)). For simplicity
the $S=2$ sector is represented in (38) by the ${\bf S}^z=+2$
states. Besides we suggest that the reduced wave function
$a_S(k,n)$ should be bounded
\begin{equation}
\sup_n{a_S(k,n)}<\infty.
\end{equation}

The ${\rm Schr\ddot odinger}$ equation for $a_S(k,n)$ has
different forms at $n>1$ and $n=1$. In the former case Eqs. (29)
and (30) give
\begin{equation}
4J_1a_S(k,n)+2J_2\cos{\frac{k}{2}}{[}a_S(k,n-1)+a_S(k,n+1){]}=Ea_S(k,n),
\end{equation}
while in the latter
\begin{equation}
(2J_1+\varepsilon_S)a_S(k,1)+2J_2\cos{\frac{k}{2}}a_S(k,2)=E(k)a_S(k,1).
\end{equation}
It is convenient to rewrite Eq. (41) in the form of Eq. (40)
\cite{13} by continuing $a_S(k,n)$ into unphysical region $n=0$.
Comparing (40) and (41) one conclude that this trick entails a
Bethe condition
\begin{equation}
\Delta_Sa_S(k,1)=\cos{\frac{k}{2}}a_S(k,0).
\end{equation}

The system (40) (considered now for $n\geq1$) and (42) together
with the restriction (39) allows us to obtain entire $a_S(k,n)$ in
a straightforward manner. Indeed representing Eq. (40) in an
equivalent matrix form
\begin{equation}
\left(\begin{array}{c}
a_S(k,n+1)\\
a_S(k,n)
\end{array}\right)
={\cal F}(\kappa)\left(\begin{array}{c}
a_S(k,n)\\
a_S(k,n-1)
\end{array}\right),
\end{equation}
where
\begin{equation}
{\cal F}(\kappa)=\left(\begin{array}{cc}
2\kappa&-1\\
1&0
\end{array}\right),\quad \kappa=\frac{E-4J_1}{4J_2\cos{k/2}},
\end{equation}
and taking $a_S(k,1):a_S(k,0)$ from (42) one consequently obtains
(up to a constant factor) using (43) the rest of $a_S(k,n)$ at
$n=2,3,\dots$ In following we shall study this problem in detail
considering separately three regions $|\kappa|<1$, $|\kappa|>1$
and $|\kappa|=1$.

For $|\kappa|\neq1$ the matrix ${\cal F}(\kappa)$ has two
different eigenvalues
\begin{equation}
\Lambda_{\pm}(\kappa)=\kappa\pm\sqrt{\kappa^2-1},
\end{equation}
related to eigenvectors
\begin{equation}
\xi_{\pm}(\kappa)=\left(\begin{array}{c}
\Lambda_{\pm}(\kappa)\\
1
\end{array}\right).
\end{equation}

At $|\kappa|<1$ it is more convenient to use the following
representation
\begin{equation}
{\Lambda_{\pm}(\kappa)}={\rm e}^{\pm iq},\quad
\kappa=\cos{q},\quad 0<q<\pi.
\end{equation}

According to (43) a decomposition
\begin{equation}
\left(\begin{array}{c}
a_S(k,1)\\
a_S(k,0)
\end{array}\right)=c_+\xi_+(\kappa)+c_-\xi_-(\kappa),
\end{equation}
($c_{\pm}$ are some coefficients) results in
\begin{equation}
\left(\begin{array}{c}
a_S(k,n+1)\\
a_S(k,n)
\end{array}\right)=\Lambda^n_+(\kappa)c_+\xi_+(\kappa)+\Lambda^n_-(\kappa)c_-\xi_-(\kappa).
\end{equation}
or equivalently
\begin{equation}
a_S^{scatt}(k,q,n)=\cos{\frac{k}{2}}\sin{qn}-\Delta_S\sin{q(n-1)}.
\end{equation}
The expression (50) obviously agrees with (39) and (as it readily
follows from (29) and (30)) corresponds to dispersion
\begin{equation}
E_{scatt}(k,q)=4\Big(J_1+J_2\cos{q}\cos{\frac{k}{2}}\Big).
\end{equation}
According to the following formulas
\begin{eqnarray}
&{\rm
e}^{ik(m+n)/2}a_S^{scatt}(k,q,n-m)=\frac{\displaystyle1}{\displaystyle2i}\Big[C_{S,12}{\rm
e}^{i(k_1m+k_2n)}-C_{S,21}{\rm e}^{i(k_2m+k_1n)}\Big],&\nonumber\\
&E_{scatt}(k,q)=E_{magn}(k_1)+E_{magn}(k_2),&
\end{eqnarray}
where
\begin{equation}
\frac{k}{2}-q=k_1<k_2=\frac{k}{2}+q,\quad q=\frac{k_2-k_1}{2},
\end{equation}
and
\begin{equation}
C_{S,ab}=\cos\frac{k_a+k_b}{2}-\Delta_S{\rm e}^{i(k_a-k_b)/2},
\end{equation}
one may associate (50) with a scattering wave function of two
magnons with wave vectors $k_1$ and $k_2$ reduced to the center
mass frame.

For $|\kappa|>1$ both the eigenvalues (45) are real. More
specifically at $\pm\kappa>1$ there should be
$|\Lambda_{\mp}(\kappa)|<1<|\Lambda_{\pm}(\kappa)|$ and the
representation (49) agree with (39) only for $c_{\pm}=0$.
According to (42), (46) and (48) in both the cases the eigenvalue
remaining in (49) is $(\cos{k/2})/\Delta_S$. So one gets
\begin{eqnarray}
a_S^{bound}(k,n)&=&\Big(\frac{\cos{k/2}}{\Delta_S}\Big)^n,\nonumber\\
E_{bound}(S,k)&=&2\Big(2J_1+J_2\Delta_S+\frac{J_2}{\Delta_S}\cos^2{\frac{k}{2}}\Big).
\end{eqnarray}
This solution exists only for
\begin{equation}
-|\Delta_S|<\cos{\frac{k}{2}}<|\Delta_S|,
\end{equation}
and according to the formulas
\begin{eqnarray}
&{\rm e}^{ik(m+n)/2}a_S^{bound}(k,n-m)={\rm
e}^{i(k_1m+k_2n)},&\nonumber\\
&E_{bound}(S,k)=E_{magn}(k_1)+E_{magn}(k_2),&
\end{eqnarray}
where
\begin{equation}
k_1=\frac{k}{2}-iv,\quad k_2=\frac{k}{2}+iv,\quad
v=\ln\Big(\frac{\Delta_S}{\cos{k/2}}\Big),
\end{equation}
may be associated with a two-magnon bound state wave function
reduced to the center mass frame. Both (52) and (57) reproduce the
Bethe Ansatze calculation presented in \cite{7}.

For $\kappa^2=1$ the matrix $\cal F(\kappa)$ has an eigenvector
$\xi_0$ and an adjoint vector $\tilde\xi_0$
\begin{equation}
\xi_0=\left(\begin{array}{c}
\kappa\\
1
\end{array}\right),\quad\tilde\xi_0=\left(\begin{array}{c}
1\\
0
\end{array}\right),
\end{equation}
for which
\begin{equation}
{\cal F}(\kappa)\xi_0(\kappa)=\kappa\xi_0(\kappa),\quad{\cal
F}(\kappa)\tilde\xi_0(\kappa)=\kappa\tilde\xi_0(\kappa)+\xi_0(\kappa).
\end{equation}
Taking into account (60) and (42) one gets the following
decomposition
\begin{equation}
\left(\begin{array}{c}
a_S(k,1)\\
a_S(k,0)
\end{array}\right)=\Delta_S\xi_0(\kappa)+(\cos{\frac{k}{2}}-\kappa\Delta_S)\tilde\xi_0(\kappa).
\end{equation}
The resulting wave function
\begin{equation}
a_S(k,\kappa,n)=n\kappa^{n-1}(\cos{\frac{k}{2}}-\kappa\Delta_S)+\kappa^n\Delta_S.
\end{equation}
agrees with (39) only on the appropriate bound of the interval
(56), namely for
\begin{equation}
\cos{\frac{k}{2}}=\kappa\Delta_S.
\end{equation}

The solution (62) may be obtained from both (50) or (55) in the
limit $|\kappa|\rightarrow1$. Indeed despite the wave function
(50) turns to zero at $q=0,\pi$ the ratio
$a^{scatt}(k,q,n)/\sin{q}$ remains finite and gives (62) as a
limit value. Analogously using the formula
$(1+\epsilon)^n=1+n\epsilon+o(\epsilon)$ one can obtain (62) from
(55).

\section{$\bf S=0$ and $\bf S=3$ three-magnon sectors}

Representing at once a $S=0$ state in general translationary
covariant form
\begin{eqnarray}
|3,0,k\rangle&=&\epsilon_{abc}\sum_{m<n<p}
{\rm e}^{ik(m+n+p)/3}b_0(k,n-m,p-n)\nonumber\\
&&\dots|1\rangle^a_m\dots|1\rangle^b_n\dots|1\rangle^c_p\dots,
\end{eqnarray}
one readily obtains from (29) and (33) the ${\rm Schr\ddot
odinger}$ equation for the reduced wave function $b_0(k,m,n)$ both
for $m,n>1$
\begin{eqnarray}
&6J_1b_0(k,m,n)+J_2[{\rm e}^{-ik/3}b_0(k,m+1,n)+{\rm
e}^{ik/3}b_0(k,m-1,n)&\nonumber\\
&+{\rm e}^{-ik/3}b_0(k,m-1,n+1)+{\rm e}^{ik/3}b_0(k,m+1,n-1)+{\rm
e}^{-ik/3}b_0(k,m,n-1)&\nonumber\\
&+{\rm e}^{ik/3}b_0(k,m,n+1)]=Eb_0(k,m,n),&
\end{eqnarray}
and $m,n=1$
\begin{eqnarray}
&(4J_1+\varepsilon_1)b_0(k,1,n)+J_2[{\rm
e}^{-ik/3}b_0(k,1,n-1)+{\rm
e}^{ik/3}b_0(k,1,n+1)&\nonumber\\
&+{\rm e}^{ik/3}b_0(k,2,n-1)+{\rm
e}^{-ik/3}b_0(k,2,n)]=Eb_0(k,1,n),&\nonumber\\
&(4J_1+\varepsilon_1)b_0(k,m,1)+J_2[{\rm
e}^{ik/3}b_0(k,m-1,1)+{\rm
e}^{-ik/3}b_0(k,m+1,1)&\nonumber\\
&+{\rm e}^{-ik/3}b_0(k,m-1,2)+{\rm
e}^{ik/3}b_0(k,m,2)]=Eb_0(k,m,1).&
\end{eqnarray}

Reduction of (66) into (65) results in a system of Bethe
conditions,
\begin{eqnarray}
2\Delta_1b_0(k,1,n)&=&{\rm e}^{ik/3}b_0(k,0,n)+{\rm e}^{-ik/3}b_0(k,0,n+1),\nonumber\\
2\Delta_1b_0(k,m,1)&=&{\rm e}^{-ik/3}b_0(k,m,0)+{\rm
e}^{ik/3}b_0(k,m+1,0).
\end{eqnarray}
The pair (65) (considered for $m,n>0$) and (67) represents the
${\rm Schr\ddot odinger}$ equation for $b_0(k,m,n)$. It is
invariant under the following duality transformation
\begin{equation}
{\cal D}(b_0(k,m,n))=\bar b_0(k,n,m).
\end{equation}
Autodual and anti-autodual solutions are related by multiplication
on $i$.

As in the two-magnon case we suggest that the reduced wave
function should be bounded
\begin{equation}
\sup_{m,n}b_0(k,m,n)<\infty.
\end{equation}

Despite the system (65), (67) is linear a proper generalization of
the straightforward matrix approach used in the previous section
is unclear for it. Instead one may treat (65) by the Fourier
substitution
\begin{equation}
\tilde b_0(k,m,n)=\varphi(k,m)\theta(k,n),
\end{equation}
which results in the following two-parametric exponential solution
\begin{equation}
\tilde b_0(k,m,n)={\rm e}^{i(\tilde q_1m+\tilde q_2n)},
\end{equation}
related to dispersion
\begin{equation}
E(k,\tilde q_1,\tilde
q_2)=2[3J_1+J_2(\cos{k_1}+\cos{k_2}+\cos{k_3})],
\end{equation}
where
\begin{equation}
k_1=\frac{k}{3}-\tilde q_1,\quad k_2=\frac{k}{3}+\tilde q_1-\tilde
q_2,\quad k_3=\frac{k}{3}+\tilde q_2.
\end{equation}

Since
\begin{equation}
{\rm e}^{ik/3(m+n+p)}\tilde b_0(k,n-m,p-n)={\rm e}^{i(
k_1m+k_2n+k_3p)},
\end{equation}
one can naturally associate (71) with the wave function of a
triple of magnons related to wave numbers $k_1$, $k_2$ and $k_3$.

Instead of $\tilde q_{1,2}$ we shall mainly use the parameters
\begin{equation}
q_1=\frac{k_2-k_1}{2}=\tilde q_1-\frac{\tilde q_2}{2},\quad
q_2=\frac{k_3-k_2}{2}=\tilde q_2-\frac{\tilde q_1}{2},
\end{equation}
considering them as generalizations of the parameter $q$ from Eq.
(53). The pair $q_{1,2}$ is more convenient for representation of
{\it pure scattering} states related to real $k_{1,2,3}$ with
$0\leq k_1<k_2<k_3\leq2\pi$ (a generalization of Eq. (2)) because
the latter system of inequalities in terms of $q_{1,2}$ has a very
simple form. Namely $0<q_{1,2}<\pi$ and $0<q_1+q_2<\pi$.
Nevertheless due to a rather compact representation (71) the
parameters $\tilde q_{1,2}$ still will be remained in exponential
factors. They also will be used for classification of states with
complex wave numbers (Eq. (82)).

When $\tilde q_{1,2}$ are complex numbers one have to treat
carefully the condition (69) and take into account that the energy
(72) must be real. These conditions result in
\begin{eqnarray}
&\Im{(\tilde q_j)}\geq0,\quad j=1,2,&\\
&\displaystyle\Big(\sin{[k/3-\Re{(\tilde
q_1})]}-\sin{[k/3+\Re{(\tilde q_1-\tilde q_2)}]}\cosh{\Im{(\tilde
q_2)}}\Big)\sinh{\Im{(\tilde
q_1)}}&\nonumber\\
&=\displaystyle\Big(\sin{[k/3+\Re{(\tilde
q_2})]}-\sin{[k/3+\Re{(\tilde q_1-\tilde q_2)}]}\cosh{\Im{(\tilde
q_1)}}\Big)\sinh{\Im{(\tilde q_2)}}.&
\end{eqnarray}

The dispersion (72) is invariant under permutations of $k_1$,
$k_2$ and $k_3$ or equivalently under the following
transformations of ${\bf\tilde q}\equiv(\tilde q_1,\tilde q_2)$
\begin{eqnarray}
\omega_1({\bf\tilde q})&=&(\tilde q_1,\tilde q_1-\tilde q_2),\quad\omega_2({\bf\tilde q})=
(-\tilde q_2,\tilde q_1-\tilde q_2),\quad\omega_3({\bf\tilde q})=(-\tilde q_2,-\tilde q_1)\nonumber\\
\omega_4({\bf\tilde q})&=&(\tilde q_2-\tilde q_1,-\tilde
q_1),\quad\omega_5({\bf\tilde q})=(\tilde q_2-\tilde q_1,\tilde
q_2).
\end{eqnarray}
In fact these formulas give a representation of the three-elements
permutation group $S_3$. It may be readily checked that all
$\omega_j$ are generated by $\omega_1$ and $\omega_5$. Namely,
\begin{equation}
\omega_2=\omega_5\cdot\omega_1,\quad\omega_3=\omega_5\cdot\omega_1\cdot\omega_5=\omega_1\cdot\omega_5\cdot\omega_1,\quad
\omega_4=\omega_1\cdot\omega_5.
\end{equation}

The symmetry (78) allows to generalize the solution (71) and
suggest the following ansatze (${\bf q}\equiv(q_1,q_2)$),
\begin{eqnarray}
b_0(k,{\bf q},m,n)&=&A_1(k,{\bf q}){\rm e}^{i(\tilde q_1m+\tilde
q_2n)}-A_2(k,{\bf q}){\rm
e}^{i(\tilde q_1m+(\tilde q_1-\tilde q_2)n)}\nonumber\\
&+&A_3(k,{\bf q}){\rm e}^{i(-\tilde q_2m+(\tilde q_1-\tilde
q_2)n)}-A_4(k,{\bf q}){\rm
e}^{-i(\tilde q_2m+\tilde q_1n)}\nonumber\\
&+&A_5(k,{\bf q}){\rm e}^{i((\tilde q_2-\tilde q_1)m-\tilde
q_1n)}-A_6(k,{\bf q}){\rm e}^{i((\tilde q_2-\tilde q_1)m+\tilde
q_2n)}.
\end{eqnarray}

For $\Im{(\tilde q_{1,2})}=0$ the expression (80) agree with (69)
while Eq. (77) is satisfied identically. However even when one of
$\tilde q_{1,2}$ has an imaginary part some of the amplitudes in
(80) must turn to zero in order to ensure an agreement with (69).
Besides according to (77) real and imaginary parts of $\tilde
q_{1,2}$ should be interdependent. More specifically let us divide
the sector $\Im{(\tilde q_{1,2})}\geq0$ on five subsectors
\begin{eqnarray}
{\cal V}_1&=&[\Im{(\tilde q_1)}>0,\, \Im{(\tilde q_2)}=0],\quad
{\cal V}_2=[\Im{(\tilde q_1)}=0,\, \Im{(\tilde
q_2)}>0],\nonumber\\
{\cal V}_3&=&[0<\Im{(\tilde q_1)}=\Im{(\tilde q_2)}],\quad {\cal
V}_4=[0<\Im{(\tilde q_1)}<\Im{(\tilde q_2)}],\nonumber\\
{\cal V}_5&=&[0<\Im{(\tilde q_2)}<\Im{(\tilde q_1)}].
\end{eqnarray}
Let ${\cal J}_i$ will be corresponding to ${\cal V}_i$ set of
$l$-s for which there should be $A_l(k,{\bf q})=0$. At the same
time ${\cal Q}_i$ will be the corresponding additional condition
on $\tilde q_{1,2}$ following from (77). For each $i$ we may
gather a triple ${\cal W}_i=[{\cal V}_i;{\cal J}_i;{\cal G}_i]$. A
straightforward analysis based on Eqs. (77) and (80) results in
the following classification
\begin{eqnarray}
{\cal W}_1&=&[\Im{(\tilde q_1)}>0,\, \Im{(\tilde
q_2)}=0;\,\{4,5,6\};
\,\Re{(\tilde q_2)}=2\Re{(\tilde q_1)}],\nonumber\\
\tilde{\cal W}_1&=&[\Im{(\tilde q_1)}>0,\, \Im{(\tilde
q_2)}=0;\,\{4,5,6\};\,\Re{(\tilde q_2)}=2k/3+\pi],\nonumber\\
{\cal W}_2&=&[\Im{(\tilde q_1)}=0,\, \Im{(\tilde
q_2)}>0;\,\{2,3,4\};
\,\Re{(\tilde q_1)}=2\Re{(\tilde q_2)}],\nonumber\\
\tilde{\cal W}_2&=&[\Im{(\tilde q_1)}=0,\, \Im{(\tilde
q_2)}>0;\,\{2,3,4\};\,\Re{(\tilde q_1)}=-2k/3+\pi],\nonumber\\
{\cal W}_3&=&[0<\Im{(\tilde q_1)}=\Im{(\tilde
q_2)};\,\{3,4,5\};\,\Re{(\tilde q_2)}=-\Re{(\tilde q_1)}],\nonumber\\
\tilde{\cal W}_3&=&[0<\Im{(\tilde q_1)}=\Im{(\tilde
q_2)};\,\{3,4,5\};\,\Re{(\tilde q_2)}=\Re{(\tilde
q_1)}+\pi-2k/3],\nonumber\\
{\cal W}_4&=&[0<\Im{(\tilde q_1)}<\Im{(\tilde q_2)};\,\{2,3,4,5\};\,{\rm Eq.}\, (77)],\nonumber\\
{\cal W}_5&=&[0<\Im{(\tilde q_2)}<\Im{(\tilde
q_1)};\,\{3,4,5,6\};\,{\rm Eq.}\, (77)].
\end{eqnarray}
Each Bethe state with complex $\tilde q$-s corresponds without
fail to one of the ${\cal W}$-s presented in (82).

The system (65) (at $m,n\geq1$), (67) exactly coincides with the
well known one inherent in the $XXZ$ model \cite{13},\cite{14}.
Nevertheless we shall give its solution within the ansatze (80) in
order to illustrate the straightforward approach used in the next
section for $S=1$ and $S=2$.

Let us begin with pure scattering states for which the duality
transformation (68) results in
\begin{equation}
{\cal D}(A_l(k,{\bf q}))=\bar A_{l-3}(k,{\bf q}),
\end{equation}
where $A_l(k,{\bf q})\equiv A_{l+6}(k,{\bf q})$ for $l=-2,-1,0$.
The $S=1$ and $S=2$ analogs of this formula are be used in
Appendix for enumeration of three-magnon Bethe states.

Substitution of (80) into (67) produces a linear system on the
amplitudes $A_l(k,{\bf q})$
\begin{equation}
\sum_{l=1}^6M^{(0)}_{il}(k,{\bf q})A_l(k,{\bf q})=0,
\end{equation}
where nonzero entries of the $6\times6$ matrix $M^{(0)}(k,{\bf
q})$ are
\begin{eqnarray}
M^{(0)}_{11}(k,{\bf q})&=&-M^{(0)}_{45}(k,{\bf q})=Z(k,\omega_5({\bf q}),\Delta_1),\nonumber\\
M^{(0)}_{22}(k,{\bf q})&=&-M^{(0)}_{56}(k,{\bf q})=Z(k,{\bf q},\Delta_1),\nonumber\\
M^{(0)}_{33}(k,{\bf q})&=&-M^{(0)}_{61}(k,{\bf q})=Z(k,\omega_1({\bf q}),\Delta_1),\nonumber\\
M^{(0)}_{44}(k,{\bf q})&=&-M^{(0)}_{12}(k,{\bf q})=Z(k,\omega_2({\bf q}),\Delta_1),\nonumber\\
M^{(0)}_{55}(k,{\bf q})&=&-M^{(0)}_{23}(k,{\bf q})=Z(k,\omega_3({\bf q}),\Delta_1),\nonumber\\
M^{(0)}_{66}(k,{\bf q})&=&-M^{(0)}_{34}(k,{\bf
q})=Z(k,\omega_4({\bf q}),\Delta_1).
\end{eqnarray}
Here
\begin{equation}
Z(k,{\bf q},\Delta)=\cos{\Big(\frac{k+q_2-q_1}{3}\Big)}-\Delta{\rm
e}^{i(q_1+q_2)},
\end{equation}
while according to (75) and (78)
\begin{eqnarray}
\omega_1({\bf q})&=&(q_1+q_2,-q_2),\quad\omega_2({\bf
q})=(-q_1-q_2,q_1),\quad\omega_3({\bf q})=(-q_2,-q_1),\nonumber\\
\omega_4({\bf q})&=&(q_2,-q_1-q_2),\quad\omega_5({\bf
q})=(-q_1,q_1+q_2).
\end{eqnarray}

Since
\begin{equation}
\det M^{(0)}(k,{\bf q} )=\prod_{n=1..6}M^{(0)}_{nn}(k,{\bf q}
)-\prod_{n=1..6}M^{(0)}_{n,n+1}(k,{\bf q})=0,
\end{equation}
(here $M^{(0)}_{67}(k,{\bf q})\equiv M^{(0)}_{61}(k,{\bf q})$) the
matrix system (84) is solvable. Namely
\begin{equation}
A_l(k,{\bf q})=\prod_{i=1}^3M^{(0)}_{l-i,l-i}(k,{\bf q}),
\end{equation}
where $M^{(0)}_{ll}(k,{\bf q})\equiv M^{(0)}_{l+6,l+6}(k,{\bf q})$
for $l=-2,-1,0$.

States with complex $\tilde q_{1,2}$ may be obtained from (89) by
analytic continuation with regard to conditions presented in the
list (82). It may be readily shown by a straightforward
calculations that there are no solutions related to $\tilde{\cal
W}_{1,2,3}$ and ${\cal W}_{4,5}$. This statement is a special
(related to the three-magnon sector) confirmation of the string
hypothesis proved for the $XXZ$ chain \cite{13},\cite{14}.

It may be readily proved that the $S=3$ case is analogous to the
$S=0$ one. It is only necessary to improve the representation (64)
(in order to obtain the state with total spin $S=3$) and replace
everywhere $\Delta_1$ on $\Delta_2$.

\section{$\bf S=1$ three-magnon sector}

General $S=1$ three-magnon state has the following representation
\begin{eqnarray}
&|3,1,k\rangle^a=\sum_{m<n<p} {\rm e}^{ik(m+n+p)/3}\Big[
b^{(1)}_1(k,n-m,p-n)&\nonumber\\
&\dots|1\rangle^a_m\dots|1\rangle^b_n\dots|1\rangle^b_p\dots+
b^{(2)}_1(k,n-m,p-n)\dots|1\rangle^b_m\dots|1\rangle^a_n\dots|1\rangle^b_p\dots&\nonumber\\
&+b^{(3)}_1(k,n-m,p-n)\dots|1\rangle^b_m\dots|1\rangle^b_n\dots|1\rangle^a_p\dots\Big],&
\end{eqnarray}
and depends on the three wave functions $b^{(1,2,3)}_1(k,m,n)$. At
$m,n>1$ the ${\rm Schr\ddot odinger}$ equation for
$b^{(1,2,3)}_1(k,m,n)$ separates on three independent linear
subsystems of the form (65) (one have only to replace $b_0(k,m,n)$
on $b^{(1,2,3)}_1(k,m,n)$). However for $m,n=1$ one gets
\begin{eqnarray}
&\Big(6J_1+J_2+\frac{\displaystyle3}{\displaystyle2}J_3\Big)b_1^{(1)}(k,1,n)+J_4b_1^{(2)}(k,1,n)+J_2\Big[{\rm
e}^{-ik/3}b_1^{(1)}(k,1,n-1)&\nonumber\\
&+{\rm e}^{ik/3}b_1^{(1)}(k,1,n+1)+{\rm
e}^{ik/3}b_1^{(1)}(k,2,n-1)&\nonumber\\
&+{\rm
e}^{-ik/3}b_1^{(1)}(k,2,n)\Big]=Eb_1^{(1)}(k,1,n),&\nonumber\\
&\Big(6J_1+J_2+\frac{\displaystyle3}{\displaystyle2}J_3\Big)b_1^{(2)}(k,1,n)+J_4b_1^{(1)}(k,1,n)+J_2\Big[{\rm
e}^{-ik/3}b_1^{(2)}(k,1,n-1)&\nonumber\\
&+{\rm e}^{ik/3}b_1^{(2)}(k,1,n+1)+{\rm
e}^{ik/3}b_1^{(2)}(k,2,n-1)&\nonumber\\
&+{\rm
e}^{-ik/3}b_1^{(2)}(k,2,n)\Big]=Eb_1^{(2)}(k,1,n),&\nonumber\\
&(4J_1+\varepsilon_0)b_1^{(3)}(k,1,n)+(J_5-J_4)(b_1^{(1)}(k,1,n)+b_1^{(2)}(k,1,n))&\nonumber\\
&+J_2[{\rm e}^{-ik/3}b_1^{(3)}(k,1,n-1)+{\rm
e}^{ik/3}b_1^{(3)}(k,1,n+1)+{\rm
e}^{ik/3}b_1^{(3)}(k,2,n-1)&\nonumber\\
&+{\rm
e}^{-ik/3}b_1^{(3)}(k,2,n)]=Eb_1^{(3)}(k,1,n),&\nonumber\\
&(4J_1+\varepsilon_0)b_1^{(1)}(k,m,1)+(J_5-J_4)[b_1^{(2)}(k,m,1)+b_1^{(3)}(k,m,1)]&\nonumber\\
&+J_2[{\rm e}^{ik/3}b_1^{(1)}(k,m-1,1)+{\rm
e}^{-ik/3}b_1^{(1)}(k,m+1,1)+{\rm
e}^{-ik/3}b_1^{(1)}(k,m-1,2)&\nonumber\\
&+{\rm
e}^{ik/3}b_1^{(1)}(k,m,2)]=Eb_1^{(1)}(k,m,1),&\nonumber\\
&\Big(6J_1+J_2+\frac{\displaystyle3}{\displaystyle2}J_3\Big)b_1^{(2)}(k,m,1)+J_4b_1^{(3)}(k,m,1)+J_2[{\rm
e}^{ik/3}b_1^{(2)}(k,m-1,1)&\nonumber\\
&+{\rm e}^{-ik/3}b_1^{(2)}(k,m+1,1)+{\rm
e}^{-ik/3}b_1^{(2)}(k,m-1,2)&\nonumber\\
&+{\rm
e}^{ik/3}b_1^{(2)}(k,m,2)]=Eb_1^{(2)}(k,m,1),&\nonumber\\
&\Big(6J_1+J_2+\frac{\displaystyle3}{\displaystyle2}J_3\Big)b_1^{(3)}(k,m,1)+J_4b_1^{(2)}(k,m,1)+J_2[{\rm
e}^{ik/3}b_1^{(3)}(k,m-1,1)&\nonumber\\
&+{\rm e}^{-ik/3}b_1^{(3)}(k,m+1,1)+{\rm
e}^{-ik/3}b_1^{(3)}(k,m-1,2)&\nonumber\\
&+{\rm e}^{ik/3}b_1^{(3)}(k,m,2)]=Eb_1^{(3)}(k,m,1).&
\end{eqnarray}

Introducing again the unphysical values $b_1^{(j)}(k,m,0)$ and
$b_1^{(j)}(k,0,n)$ one can reduce (91) to the form (65) by
producing the following system of Bethe conditions
\begin{eqnarray}
&(\Delta_2+\Delta_1)b^{(1)}_1(k,1,n)+(\Delta_2-\Delta_1)b^{(2)}_1(k,1,n)=
{\rm e}^{ik/3}b^{(1)}_1(k,0,n)&\nonumber\\
&+{\rm e}^{-ik/3}b^{(1)}_1(k,0,n+1),&\nonumber\\
&(\Delta_2+\Delta_1)b^{(2)}_1(k,1,n)+(\Delta_2-\Delta_1)b^{(1)}_1(k,1,n)=
{\rm e}^{ik/3}b^{(2)}_1(k,0,n)&\nonumber\\
&+{\rm e}^{-ik/3}b^{(2)}_1(k,0,n+1),&\nonumber\\
&2\Delta_0b^{(3)}_1(k,1,n)+\frac{2}{3}(\Delta_0-\Delta_2)[b^{(1)}_1(k,1,n)+b^{(2)}_1(k,1,n)]&\nonumber\\
&={\rm e}^{ik/3}b^{(3)}_1(k,0,n)
+{\rm e}^{-ik/3}b^{(3)}_1(k,0,n+1),&\nonumber\\
&2\Delta_0b^{(1)}_1(k,m,1)+\frac{2}{3}(\Delta_0-\Delta_2)[b^{(2)}_1(k,m,1)+b^{(3)}_1(k,m,1)]&\nonumber\\
&={\rm e}^{-ik/3}b^{(1)}_1(k,m,0)+{\rm
e}^{ik/3}b^{(1)}_1(k,m+1,0),&\nonumber\\
&(\Delta_2+\Delta_1)b^{(2)}_1(k,m,1)+(\Delta_2-\Delta_1)b^{(3)}_1(k,m,1)=
{\rm e}^{-ik/3}b^{(2)}_1(k,m,0)&\nonumber\\
&+{\rm e}^{ik/3}b^{(2)}_1(k,m+1,0),&\nonumber\\
&(\Delta_2+\Delta_1)b^{(3)}_1(k,m,1)+(\Delta_2-\Delta_1)b^{(2)}_1(k,m,1)=
{\rm e}^{-ik/3}b^{(3)}_1(k,m,0)&\nonumber\\
&+{\rm e}^{ik/3}b^{(3)}_1(k,m+1,0).&
\end{eqnarray}

As it was in the $S=0$ case the system (92) (as well as the three
separate subsystems of the form (65) for $b^{(j)}_1(k,m,n)$) is
symmetric under a duality transformation
\begin{equation}
{\cal D}(b^{(j)}_1(k,m,n))=\bar b^{(4-j)}_1(k,n,m).
\end{equation}

Before developing a general analysis of the system (92) we shall
at once find all the cases when it may be reduced to the
$XXZ$-type form (67).

First of all for
\begin{equation}
\Delta_0=\Delta_1=\Delta_2
\end{equation}
the system (92) decouples into three $XXZ$-type subsystems (67)
and therefore is completely solvable. Namely
\begin{equation}
b_1^{(j)}(k,{\bf q},m,n)=\alpha_jb_0(k,{\bf q},m,n), \quad
j=1,2,3,
\end{equation}
where $\alpha_{1,2,3}$ is a triple of arbitrary parameters.
Labeling the coupling constants related to (94) by an upper index
"$(0)$" one may readily obtain from (32)
\begin{equation}
J^{(0)}_d=-J^{(0)}_l,\quad
J^{(0)}_{ll}=4J^{(0)}_l,\quad\Delta^{(0)}_{0,1,2}=\frac{J^{(0)}_{rr}}{4J^{(0)}_l},
\end{equation}
or equivalently
\begin{equation}
J_4^{(0)}=J_5^{(0)}=0,\qquad\Delta^{(0)}_{0,1,2}=1+\frac{3J_3^{(0)}}{2J_2^{(0)}}.
\end{equation}
In this case according to (17) and (97) an interaction between
excited triplet rungs is spin-independent. In more detail the
relation between this model and $XXZ$ chain was studied in
\cite{27}.

Besides the complete separable case (94) there are also two
configurations of $\Delta$-s for which the system (92) possess a
partial solution of the form (95) however with special values of
the ratios $\alpha_i/\alpha_j$ ($i,j=1,2,3$). Indeed substituting
the ansatze (95) into the system (92) one readily makes sure that
the latter may be reduced to (67) with an appropriate parameter
$\Delta$ only under the following system of conditions
\begin{eqnarray}
(\Delta_1+\Delta_2-2\Delta)\alpha_1+(\Delta_2-\Delta_1)\alpha_2&=&0,\nonumber\\
(\Delta_2-\Delta_1)\alpha_1+(\Delta_1+\Delta_2-2\Delta)\alpha_2&=&0,\\
(\Delta_1+\Delta_2-2\Delta)\alpha_2+(\Delta_2-\Delta_1)\alpha_3&=&0,\nonumber\\
(\Delta_2-\Delta_1)\alpha_2+(\Delta_1+\Delta_2-2\Delta)\alpha_3&=&0,\\
3(\Delta_0-\Delta)\alpha_1+(\Delta_0-\Delta_2)(\alpha_2+\alpha_3)&=&0,\nonumber\\
3(\Delta_0-\Delta)\alpha_3+(\Delta_0-\Delta_2)(\alpha_1+\alpha_2)&=&0.
\end{eqnarray}

It may be easily observed that a trivial solution of the subsystem
(98) may be nontrivially extended as a solution of the whole
system (98)-(100) only in the case (94). From the other hand a
nontrivial solution of (98), namely $\alpha_1=\alpha_2$, exists
only for $\Delta=\Delta_2$. Besides for $\Delta_1=\Delta_2=\Delta$
the system (98) is satisfied for all $\alpha_{1,2}$. Extension of
these two solutions on the subsystems (99), (100) results in
\begin{eqnarray}
\alpha_1=\alpha_2=\alpha_3,&\quad&
\Delta=\Delta_0=\Delta_2,\\
4\alpha_1=-\alpha_2=4\alpha_3,&\quad& \Delta=\Delta_1=\Delta_2.
\end{eqnarray}

Turning to the general ($XXZ$-irreducible) case we suggest the
Bethe ansatze,
\begin{eqnarray}
&b^{(j)}_1(k,m,n)=B^{(j)}_1(k,{\bf q}){\rm e}^{i(\tilde
q_1m+\tilde q_2n)}-B^{(j)}_2(k,{\bf q}){\rm
e}^{i(\tilde q_1m+(\tilde q_1-\tilde q_2)n)}&\nonumber\\
&+B^{(j)}_3(k,{\bf q}){\rm e}^{i(-\tilde q_2m+(\tilde q_1-\tilde
q_2)n)}-B^{(j)}_4(k,{\bf q}){\rm
e}^{-i(\tilde q_2m+\tilde q_1n)}&\nonumber\\
&+B^{(j)}_5(k,{\bf q}){\rm e}^{i((\tilde q_2-\tilde q_1)m-\tilde
q_1n)}-B^{(j)}_6(k,{\bf q}){\rm e}^{i((\tilde q_2-\tilde
q_1)m+\tilde q_2n)}.&
\end{eqnarray}

Classification of states with complex $\tilde q_{1,2}$ has the
form (82). However each ${\cal J}_i$ in (82) is now a set of $l$-s
for which all $B^{(j)}_l(k,{\bf q})=0$. In the present paper we
shall not study $S=1$ and $S=2$ three-magnon Bethe states with
complex wave numbers. For the pure scattering states the duality
(93) reduces on the amplitudes as follows
\begin{equation}
{\cal D}(B^{(j)}_l(k,{\bf q}))=\bar B^{(j)}_{l-3}(k,{\bf q}),
\end{equation}

Substitution of (103) into (92) gives
\begin{equation}
\sum_{l=1}^{18}M^{(1)}_{il}(k,{\bf q})B_l(k,{\bf q})=0,
\end{equation}
where the vector column $B_l(k,{\bf q})$ for $l=1,...,18$ is
defined as follows
\begin{equation}
B_{6(j-1)+m}(k,{\bf q})=B_m^{(j)}(k,{\bf q}),\quad j=1,2,3,\quad
m=1,...,6,
\end{equation}
while nonzero entries of the $18\times18$ matrix $M^{(1)}(k,{\bf
q})$ are the following,
\begin{eqnarray}
M^{(1)}_{11}(k,{\bf q})&=&-M^{(1)}_{16,17}(k,{\bf q})=Z(k,\omega_5({\bf q}),\Delta_0),\nonumber\\
M^{(1)}_{22}(k,{\bf q})&=&M^{(1)}_{88}(k,{\bf
q})=-M^{(1)}_{11,12}(k,{\bf q})=
-M^{(1)}_{17,18}(k,{\bf q})=Z\Big(k,{\bf q},\frac{\Delta_1+\Delta_2}{2}\Big),\nonumber\\
M^{(1)}_{33}(k,{\bf q})&=&-M^{(1)}_{18,13}(k,{\bf q})=Z(k,\omega_1({\bf q}),\Delta_0),\nonumber\\
M^{(1)}_{44}(k,{\bf q})&=&M^{(1)}_{10,10}(k,{\bf
q})=-M^{(1)}_{78}(k,{\bf q})=-M^{(1)}_{13,14}(k,{\bf q})=
Z\Big(k,\omega_2({\bf q}),\frac{\Delta_1+\Delta_2}{2}\Big),\nonumber\\
M^{(1)}_{55}(k,{\bf q})&=&-M^{(1)}_{14,15}(k,{\bf q})=Z(k,\omega_3({\bf q}),\Delta_0\Big),\nonumber\\
M^{(1)}_{66}(k,{\bf q})&=&M^{(1)}_{12,12}(k,{\bf
q})=-M^{(1)}_{9,10}(k,{\bf q})=-M^{(1)}_{15,16}(k,{\bf q})=
Z\Big(k,\omega_4({\bf q}),\frac{\Delta_1+\Delta_2}{2}\Big),\nonumber\\
M^{(1)}_{77}(k,{\bf q})&=&M^{(1)}_{13,13}(k,{\bf
q})=-M^{(1)}_{45}(k,{\bf q})=-M^{(1)}_{10,11}(k,{\bf q})=
Z\Big(k,\omega_5({\bf q}),\frac{\Delta_1+\Delta_2}{2}\Big),\nonumber\\
M^{(1)}_{99}(k,{\bf q})&=&M^{(1)}_{15,15}(k,{\bf
q})=-M^{(1)}_{61}(k,{\bf q})=-M^{(1)}_{12,7}(k,{\bf q})=
Z\Big(k,\omega_1({\bf q}),\frac{\Delta_1+\Delta_2}{2}\Big),\nonumber\\
M^{(1)}_{11,11}(k,{\bf q})&=&M^{(1)}_{17,17}(k,{\bf
q})=-M^{(1)}_{23}(k,{\bf q})=-M^{(1)}_{89}(k,{\bf q})=
Z\Big(k,\omega_3({\bf q}),\frac{\Delta_1+\Delta_2}{2}\Big),\nonumber\\
M^{(1)}_{14,14}(k,{\bf q})&=&-M^{(1)}_{56}(k,{\bf q})=Z({\bf q},\Delta_0),\nonumber\\
M^{(1)}_{16,16}(k,{\bf q})&=&-M^{(1)}_{12}(k,{\bf q})=Z(k,\omega_2({\bf q}),\Delta_0),\nonumber\\
M^{(1)}_{18,18}(k,{\bf q})&=&-M^{(1)}_{34}(k,{\bf q})=Z(k,\omega_4({\bf q}),\Delta_0),\nonumber\\
M^{(1)}_{17}(k,{\bf q})&=&M^{(1)}_{1,13}(k,{\bf
q})=-M^{(1)}_{16,5}(k,{\bf q})=-M^{(1)}_{16,11}(k,{\bf
q})=\frac{\Delta_2-\Delta_0}{3}{\rm
e}^{iq_2},\nonumber\\
M^{(1)}_{18}(k,{\bf q})&=&M^{(1)}_{1,14}(k,{\bf
q})=-M^{(1)}_{16,4}(k,{\bf q})=-M^{(1)}_{16,10}(k,{\bf
q})=\frac{\Delta_0-\Delta_2}{3}{\rm
e}^{-iq_2},\nonumber\\
M^{(1)}_{39}(k,{\bf q})&=&M^{(1)}_{3,15}(k,{\bf
q})=-M^{(1)}_{18,1}(k,{\bf q})=-M^{(1)}_{18,7}(k,{\bf
q})=\frac{\Delta_2-\Delta_0}{3}{\rm
e}^{iq_1},\nonumber\\
M^{(1)}_{3,10}(k,{\bf q})&=&M^{(1)}_{3,16}(k,{\bf
q})=-M^{(1)}_{18,6}(k,{\bf q})=-M^{(1)}_{18,12}(k,{\bf
q})=\frac{\Delta_0-\Delta_2}{3}{\rm
e}^{-iq_1},\nonumber\\
M^{(1)}_{5,11}(k,{\bf q})&=&M^{(1)}_{5,17}(k,{\bf
q})=-M^{(1)}_{14,3}(k,{\bf q})=-M^{(1)}_{14,9}(k,{\bf
q})=\frac{\Delta_2-\Delta_0}{3}{\rm
e}^{-i(q_1+q_2)},\nonumber\\
M^{(1)}_{5,12}(k,{\bf q})&=&M^{(1)}_{5,18}(k,{\bf
q})=-M^{(1)}_{14,2}(k,{\bf q})=-M^{(1)}_{14,8}(k,{\bf
q})=\frac{\Delta_0-\Delta_2}{3}{\rm
e}^{i(q_1+q_2)},\nonumber\\
M^{(1)}_{28}(k,{\bf q})&=&M^{(1)}_{82}(k,{\bf
q})=-M^{(1)}_{11,18}(k,{\bf q})=-M^{(1)}_{17,12}(k,{\bf
q})=\frac{\Delta_1-\Delta_2}{2}{\rm
e}^{i(q_1+q_2)},\nonumber\\
M^{(1)}_{29}(k,{\bf q})&=&M^{(1)}_{83}(k,{\bf
q})=-M^{(1)}_{11,17}(k,{\bf q})=-M^{(1)}_{17,11}(k,{\bf
q})=\frac{\Delta_2-\Delta_1}{2}{\rm
e}^{-i(q_1+q_2)},\nonumber\\
M^{(1)}_{4,10}(k,{\bf q})&=&M^{(1)}_{10,4}(k,{\bf
q})=-M^{(1)}_{7,14}(k,{\bf q})=-M^{(1)}_{13,8}(k,{\bf
q})=\frac{\Delta_1-\Delta_2}{2}{\rm
e}^{-iq_2},\nonumber\\
M^{(1)}_{4,11}(k,{\bf q})&=&M^{(1)}_{10,5}(k,{\bf
q})=-M^{(1)}_{7,13}(k,{\bf q})=-M^{(1)}_{13,7}(k,{\bf
q})=\frac{\Delta_2-\Delta_1}{2}{\rm
e}^{iq_2},\nonumber\\
M^{(1)}_{67}(k,{\bf q})&=&M^{(1)}_{12,1}(k,{\bf
q})=-M^{(1)}_{9,15}(k,{\bf q})=-M^{(1)}_{15,9}(k,{\bf
q})=\frac{\Delta_2-\Delta_1}{2}{\rm
e}^{iq_1},\nonumber\\
M^{(1)}_{6,12}(k,{\bf q})&=&M^{(1)}_{12,6}(k,{\bf
q})=-M^{(1)}_{9,16}(k,{\bf q})=-M^{(1)}_{15,10}(k,{\bf
q})=\frac{\Delta_1-\Delta_2}{2}{\rm e}^{-iq_1}.
\end{eqnarray}

As in the case (94) for complete solvability of the $S=1$ problem
it is necessary to have three independent solutions of the system
(92). In the Bethe Ansatze framework (103), (105) this results in
\begin{equation}
{\rm rank}(M^{(1)}(k,{\bf q}))=15,
\end{equation}
and therefore in
\begin{equation}
P^{(1)}_n(k,{\bf q})=0,\quad n=0,1,2.
\end{equation}
Here $P^{(1)}_n(k,{\bf q})$ are coefficients of the characteristic
polynom
\begin{equation}
|M^{(1)}(k,{\bf q})-\lambda I|=\sum_{n=0}^{18}P_n^{(1)}(k,{\bf
q})\lambda^n.
\end{equation}

Direct calculation based on the computer algebra system MAPLE
gives
\begin{equation}
P_0^{(1)}(k,{\bf q})={\det}M^{(1)}(k,{\bf q})=0,
\end{equation}
therefore even in the general case ${\rm rank}(M^{(1)}(k,{\bf
q}))\leq17$ and the system (105) {\it always} has at the minimum
one solution. Its general form is represented in the Appendix A.

For the next coefficient $P^{(1)}_1(k,{\bf q})$ we have obtained
by machinery calculations the following factorization
\begin{equation}
P_1^{(1)}(k,{\bf q})=\frac{2}{729}\Big(1-{\rm
e}^{2iq_1}\Big)^2\Big(1-{\rm e}^{2iq_2}\Big)^2\Big(1-{\rm
e}^{2i(q_1+q_2)}\Big)^2\tilde P_1^{(1)}(k,{\bf q}),
\end{equation}
where
\begin{equation}
\tilde P_1^{(1)}(k,{\bf q})={\rm e}^{-i(11k+31q_1+31q_2)/3}
\sum_{m,n,p\geq0}Q_{m,n,p}(\Delta_0,\Delta_1,\Delta_2){\rm
e}^{i(mk+nq_1+pq_2)/3}.
\end{equation}
The sum in (113) contains 95052 terms (that is why $\tilde
P_1^{(1)}(k,{\bf q})$ can not be represented in the format of this
paper). According to (112) and (75) the condition (109) is
satisfied at $n=1$ if either any two wave numbers from $k_1$,
$k_2$ and $k_3$ coincides or at
\begin{equation}
\tilde P_1^{(1)}(k,{\bf q})=0.
\end{equation}
The former three cases are similar to the case $k_1=k_2$ in the
two-magnon problem studied in the Section 3. Three-magnon
solutions of this type will not studied in the present paper.
Turning to the Eq. (114) we shall confine ourselves by the problem
of its solvability for all wave numbers. Namely we shall postulate
\begin{equation}
Q_{m,n,p}(\Delta_0,\Delta_1,\Delta_2)=0,
\end{equation}
to be valid at all $m,n,p$.

Despite the system (115) depends only on a triple of unknown
variables it is practically unsolvable by the MAPLE ${\rm Gr\ddot
obner}$ package on a personal computer with RAM about 2 Gb.
Luckily (as it may be directly checked by machinery calculation)
\begin{equation}
\frac{2Q_{8,4,13}-9Q_{12,0,11}-Q_{4,8,15}}{2592\Delta^2_1\Delta^2_2\Delta^2_3}=12(\Delta_0-\Delta_1)^2+
15(\Delta_1-\Delta_2)^2+20(\Delta_2-\Delta_0)^2,
\end{equation}
so except (94) there are no solutions with
$\Delta_0\Delta_1\Delta_2\neq0$.

In each of the three cases $\Delta_{0,1,2}=0$ the reduced system
(115) is essentially simpler than the initial one and may be
readily solved on the personal computer. Calculations based on the
${\rm Gr\ddot obner}$ package gave four pairs of solutions. We
shall represent them as sets of $\Delta$-parameters:
${\bf\Delta}=[\Delta_0,\Delta_1,\Delta_2]$, and additionally as
the corresponding sets of the coupling constants: ${\bf
J}=[J_l,J_d,J_{rr},J_{ll},J_{dd}]$ and $\tilde{\bf
J}=[J_2,J_3,J_4,J_5]$. Note that the parameters $J_r$ and $J_1$
remain indefinite. This is rather evident because both of them
correspond to the term proportional to $\hat Q$ in the
Hamiltonian. But according to (23) the former has no affect on the
Bethe equations. Namely the solutions are the following
\begin{eqnarray}
{\bf\Delta}^{(1,\pm)}&=&[\pm1,0,\pm1],\quad{\bf J}^{(1,+)}=[1,0,0,4,0],\quad\tilde{\bf J}^{(1,\pm)}=[\pm1,0,1,1],\\
{\bf\Delta}^{(2,\pm)}&=&[0,\pm1,0],\quad{\bf
J}^{(2,+)}=[1,0,-2,4,0],\quad\tilde{\bf
J}^{(2,\pm)}=[\pm1,-2,1,1],\\
{\bf\Delta}^{(3,\pm)}&=&[0,\pm\frac{3}{2},\pm\frac{3}{2}],\quad{\bf J}^{(3,+)}=[1,0,4,0,-4],\quad
\tilde{\bf J}^{(3,\pm)}=[\pm1,4,0,-1],\\
{\bf\Delta}^{(4,\pm)}&=&[\mp\frac{3}{2},0,0],\quad{\bf
J}^{(4,+)}=[1,0,1,0,-4],\quad\tilde{\bf
J}^{(4,\pm)}=[\pm1,1,0,-1].
\end{eqnarray}
It may be readily shown that the condition (27) is satisfied only
for "+" solutions, while the "-" ones may be obtained from them by
the symmetry (28). That is why we have omitted representations for
${\bf J}^{(1,2,3,4,-)}$. However they may be readily obtained from
$\tilde{\bf J}^{(1,2,3,4,-)}$ using Eqs. (20).

The models related to ${\bf\Delta}^{1,+}$ and ${\bf\Delta}^{2,+}$
were first presented in \cite{28}. Then the former one was
intensively studied in \cite{3}. Algebraic structures related to
${\bf\Delta}^{1,+}$ and ${\bf\Delta}^{3,-}$ models as well as to
the model (94) with $\Delta_{0,1,2}=1$ were presented in
\cite{23}. (However the cases (118), (120) and the general case
(94) were not discussed in \cite{23}).

As it follows from (101) and (102) all the models (117)-(120) have
a $XXZ$-type solution (95). The remaining pair of solutions may be
chosen in different ways. (In other words we do not know the best
choice of basis in the two-dimensional solution subspace
additional to (95)). The basises obtained by machinery
calculations within MAPLE are presented in the Appendix.

\section{$\bf S=2$ three-magnon sector}

A $S=2$ three magnon state related to ${\bf S}^z=2$ has the
following form
\begin{eqnarray}
|3,2,k\rangle^a&=&\sum_{m<n<p}
{\rm e}^{ik(m+n+p)/3}[\nonumber\\
&&b^{(1)}_2(k,n-m,p-n)\dots|1\rangle^+_m\dots(|1\rangle^+_n\dots|1\rangle^3_p-|1\rangle^3_n\dots|1\rangle^+_p)...\nonumber\\
&+&
b^{(2)}_2(k,n-m,p-n)\dots(|1\rangle^+_m\dots|1\rangle^3_n-|1\rangle^3_m\dots|1\rangle^+_n)\dots|1\rangle^+_p\dots
\end{eqnarray}

For $m,n>1$ the amplitudes $b^{(1,2)}_2(k,m,n)$ separately satisfy
the ${\rm Schr\ddot odinger}$ equation (65) while for $m=1$ or
$n=1$
\begin{eqnarray}
&(4J_1+\varepsilon_2)b_2^{(1)}(k,1,n)+J_2[{\rm
e}^{-ik/3}b_2^{(1)}(k,1,n-1)+{\rm e}^{ik/3}b_2^{(1)}(k,1,n+1)&\nonumber\\
&+{\rm e}^{ik/3}b_2^{(1)}(k,2,n-1)+{\rm
e}^{-ik/3}b_2^{(1)}(k,2,n)]=Eb_2^{(1)}(k,1,n),&\nonumber\\
&(4J_1+\varepsilon_1)b_2^{(2)}(k,1,n)+J_4b_2^{(1)}(k,1,n)+J_2[{\rm
e}^{-ik/3}b_2^{(2)}(k,1,n-1)&\nonumber\\
&+{\rm e}^{ik/3}b_2^{(2)}(k,1,n+1)+{\rm
e}^{ik/3}b_2^{(2)}(k,2,n-1)&\nonumber\\
&+{\rm
e}^{-ik/3}b_2^{(2)}(k,2,n)]=Eb_2^{(2)}(k,1,n),&\nonumber\\
&(4J_1+\varepsilon_2)b_2^{(2)}(k,m,1)+J_2[{\rm
e}^{ik/3}b_2^{(2)}(k,m-1,1)+{\rm e}^{-ik/3}b_2^{(2)}(k,m+1,1)&\nonumber\\
&+{\rm e}^{-ik/3}b_2^{(2)}(k,m-1,2)+{\rm
e}^{ik/3}b_2^{(2)}(k,m,2)]=Eb_2^{(2)}(k,m,1),&\nonumber\\
&(4J_1+\varepsilon_1)b_2^{(1)}(k,m,1)+J_4b_2^{(2)}(k,m,1)+J_2[{\rm
e}^{ik/3}b_2^{(1)}(k,m-1,1)&\nonumber\\
&+{\rm e}^{-ik/3}b_2^{(1)}(k,m+1,1)+{\rm
e}^{-ik/3}b_2^{(1)}(k,m-1,2)&\nonumber\\
&+{\rm e}^{ik/3}b_2^{(1)}(k,m,2)]=Eb_2^{(1)}(k,m,1).&
\end{eqnarray}
Introducing again the unphysical amplitudes we obtain from (122)
the corresponding system of coupled Bethe conditions
\begin{eqnarray}
&2\Delta_1b_2^{(1)}(k,m,1)+(\Delta_2-\Delta_1)b_2^{(2)}(k,m,1)={\rm
e}^{-ik/3}b_2^{(1)}(k,m,0)+{\rm
e}^{ik/3}b_2^{(1)}(k,m+1,0),&\nonumber\\
&2\Delta_2b_2^{(2)}(k,m,1)={\rm e}^{-ik/3}b_2^{(2)}(k,m,0)+{\rm
e}^{ik/3}b_2^{(2)}(k,m+1,0),&\nonumber\\
&2\Delta_1b_2^{(2)}(k,1,n)+(\Delta_2-\Delta_1)b_2^{(1)}(k,1,n)={\rm
e}^{ik/3}b_2^{(2)}(k,0,n)+{\rm
e}^{-ik/3}b_2^{(2)}(k,0,n+1),&\nonumber\\
&2\Delta_2b_2^{(1)}(k,1,n)={\rm e}^{ik/3}b_2^{(1)}(k,0,n)+{\rm
e}^{-ik/3}b_2^{(1)}(k,0,n+1),&
\end{eqnarray}
invariant under duality transformation
\begin{equation}
{\cal D}(b^{(j)}_2(k,m,n))=\bar b^{(3-j)}_2(k,n,m).
\end{equation}

For
\begin{equation}
\Delta_1=\Delta_2,
\end{equation}
or (according to (32))
\begin{equation}
J_{ll}=-4J_d,
\end{equation}
this system decouples into a pair of the $XXZ$-type subsystems
(67) on $b_2^{(1,2)}(k,m,n)$. In this case the general solution
\begin{equation}
b_2^{(j)}(k,{\bf q},m,n)=\beta_jb_0(k,{\bf q},m,n),\quad j=1,2,
\end{equation}
depends on ${\bf q}$ and two arbitrary parameters $\beta_{1,2}$.

It may be readily proved that the $XXZ$-type solutions (127) exist
only under the condition (125).

In the general case making the standard substitution
\begin{eqnarray}
&b^{(j)}_2(k,m,n)=C^{(j)}_1(k,{\bf q}){\rm e}^{i(\tilde
q_1m+\tilde q_2n)}-C^{(j)}_2(k,{\bf q}){\rm
e}^{i(\tilde q_1m+(\tilde q_1-\tilde q_2)n)}&\nonumber\\
&+C^{(j)}_3(k,{\bf q}){\rm e}^{i(-\tilde q_2m+(\tilde q_1-\tilde
q_2)n)}-C^{(j)}_4(k,{\bf q}){\rm
e}^{-i(\tilde q_2m+\tilde q_1n)}&\nonumber\\
&+C^{(j)}_5(k,{\bf q}){\rm e}^{i((\tilde q_2-\tilde q_1)m-\tilde
q_1n)}-C^{(j)}_6(k,{\bf q}){\rm e}^{i((\tilde q_2-\tilde
q_1)m+\tilde q_2n)},&
\end{eqnarray}
one results in a linear system
\begin{equation}
\sum_{j=1}^{12}M^{(2)}_{ij}(k,{\bf q})C_j(k,{\bf q})=0,
\end{equation}
where as in (106)
\begin{equation}
C_{6(j-1)+m}(k,{\bf q})=C_m^{(j)}(k,{\bf q}),\quad j=1,2,\quad
m=1,...,6,
\end{equation}
and the $12\times12$ matrix $M^{(2)}(k,{\bf q})$ has the following
nonzero entries
\begin{eqnarray}
M^{(2)}_{11}(k,{\bf q})&=&-M^{(2)}_{10,11}(k,{\bf q})=Z(k,\omega_5({\bf q}),\Delta_1),\nonumber\\
M^{(2)}_{22}(k,{\bf q})&=&-M^{(2)}_{11,12}(k,{\bf q})=Z(k,{\bf q},\Delta_2),\nonumber\\
M^{(2)}_{33}(k,{\bf q})&=&-M^{(2)}_{12,7}(k,{\bf q})=Z(k,\omega_1({\bf q}),\Delta_1),\nonumber\\
M^{(2)}_{44}(k,{\bf q})&=&-M^{(2)}_{78}(k,{\bf q})=Z(k,\omega_2({\bf q}),\Delta_2),\nonumber\\
M^{(2)}_{55}(k,{\bf q})&=&-M^{(2)}_{89}(k,{\bf q})=Z(k,\omega_3({\bf q}),\Delta_1),\nonumber\\
M^{(2)}_{66}(k,{\bf q})&=&-M^{(2)}_{9,10}(k,{\bf q})=Z(k,\omega_4({\bf q}),\Delta_2),\nonumber\\
M^{(2)}_{77}(k,{\bf q})&=&-M^{(2)}_{45}(k,{\bf q})=Z(k,\omega_5({\bf q}),\Delta_2),\nonumber\\
M^{(2)}_{88}(k,{\bf q})&=&-M^{(2)}_{56}(k,{\bf q})=Z(k,{\bf q},\Delta_1),\nonumber\\
M^{(2)}_{99}(k,{\bf q})&=&-M^{(2)}_{61}(k,{\bf q})=Z(k,\omega_1({\bf q}),\Delta_2),\nonumber\\
M^{(2)}_{10,10}(k,{\bf q})&=&-M^{(2)}_{12}(k,{\bf q})=Z(k,\omega_2({\bf q}),\Delta_1),\nonumber\\
M^{(2)}_{11,11}(k,{\bf q})&=&-M^{(2)}_{23}(k,{\bf q})=Z(k,\omega_3({\bf q}),\Delta_2),\nonumber\\
M^{(2)}_{12,12}(k,{\bf q})&=&-M^{(2)}_{34}(k,{\bf q})=Z(k,\omega_4({\bf q}),\Delta_1),\nonumber\\
M^{(2)}_{17}(k,{\bf q})&=&-M^{(2)}_{10,5}(k,{\bf
q})=\frac{\Delta_1-\Delta_2}{2}{\rm
e}^{iq_2},\nonumber\\
M^{(2)}_{18}(k,{\bf q})&=&-M^{(2)}_{10,4}(k,{\bf
q})=\frac{\Delta_2-\Delta_1}{2}{\rm
e}^{-iq_2},\nonumber\\
M^{(2)}_{39}(k,{\bf q})&=&-M^{(2)}_{12,1}(k,{\bf
q})=\frac{\Delta_1-\Delta_2}{2}{\rm
e}^{iq_1},\nonumber\\
M^{(2)}_{3,10}(k,{\bf q})&=&-M^{(2)}_{12,6}(k,{\bf
q})=\frac{\Delta_2-\Delta_1}{2}{\rm
e}^{-iq_1},\nonumber\\
M^{(2)}_{5,11}(k,{\bf q})&=&-M^{(2)}_{83}(k,{\bf
q})=\frac{\Delta_1-\Delta_2}{2}{\rm
e}^{-i(q_1+q_2)},\nonumber\\
M^{(2)}_{5,12}(k,{\bf q})&=&-M^{(2)}_{82}(k,{\bf
q})=\frac{\Delta_2-\Delta_1}{2}{\rm e}^{i(q_1+q_2)}.
\end{eqnarray}

As in the $S=1$ case we shall concern only on the pure scattering
states for which the duality (124) gives
\begin{equation}
{\cal D}(C^{(j)}_l(k,{\bf q}))=\bar C^{(j)}_{l-3}(k,{\bf q}).
\end{equation}

To be completely solvable the system (129) must posses two
independent solutions. Equivalently there should be
\begin{equation}
{\rm rank}(M^{(2)}(k,{\bf q}))=10.
\end{equation}
According to machinery calculation
\begin{equation}
{\rm det}M^{(2)}(k,{\bf q})=-6(\Delta_1-\Delta_2)^2(1-{\rm
e}^{2i(q_1+q_2)})^2 (1-{\rm e}^{-2iq_1})^2(1-{\rm
e}^{-2iq_2})^2Y^2(k,{\bf q}),
\end{equation}
where
\begin{eqnarray}
&\displaystyle Y(k,{\bf q})=
\displaystyle\Big[(\Delta_1-\Delta_2)^2-1\Big]\cos{k}
+\Big[(\Delta_1+\Delta_2)^2-1\Big]\Big[\cos{\frac{k-4q_1-2q_2}{3}}&\nonumber\\
&\displaystyle+\cos{\frac{k+2q_1+4q_2}{3}}+\cos{\frac{k+2q_1-2q_2}{3}}\Big]-4\Delta_1\Delta_2(\Delta_1+\Delta_2).&
\end{eqnarray}

A condition
\begin{equation}
{\det}M^{(2)}(k,{\bf q})=0,
\end{equation}
will be satisfied at all $k$, $q_1$ and $q_2$ either in the case
(125) or in the four additional ones
\begin{equation}
\Delta_1=\pm1,\quad\Delta_2=0
\end{equation}
and
\begin{equation}
\Delta_1=0,\quad\Delta_2=\pm1.
\end{equation}
Machinery calculations show that in all these cases the condition
(133) is satisfied. The corresponding solutions of the system
(129) are presented in the Appendix.

\section{Integrability and the Reshetikhin condition}

A well known alternative to the Coordinate Bethe Ansatze is the so
called Algebraic Bethe Ansatze or the Inverse Scattering Method
\cite{15}-\cite{17}. It is based on the representation of the
finite dimensional matrix $H$ related to the local Hamiltonian
density $H_{n,n+1}$ as a derivative of the corresponding
$R$-matrix.
\begin{equation}
H=\frac{\partial}{\partial\lambda}\check
 R(\lambda)|_{\lambda=0}.
\end{equation}
The latter satisfies the Yang-Baxter equation,
\begin{equation}
\check R_{12}(\la-\mu)\check R_{23}(\la)\check R_{12}(\mu)=\check
R_{23}(\mu)\check R_{12}(\la)\check R_{23}(\la-\mu),
\end{equation}
and the initial condition,
\begin{equation}
\check R(0)\propto I,
\end{equation}
(where again $I$ is an identity matrix).

From (139)-(141) follows the Reshetikhin condition \cite{17},
\begin{equation}
{[}H_{12}+H_{23},{[}H_{12},H_{23}{]]}=K_{23}-K_{12},
\end{equation}
which for the Hamiltonian density (17) with $J_6=0$ gives the
following system of equations,
\begin{eqnarray}
J_2J_4(J_1+J_3+J_5)&=&0,\nonumber\\
J_2(J_4-J_5)(J_4+J_5)&=&0,\nonumber\\
J_2(J_4-J_5)(2J_1+2J_3-J_4+5J_5)&=&0,\nonumber\\
(J_4-J_5)(J_2^2-J_5^2+2J_4J_5)&=&0,\nonumber\\
J_5(J_2^2-2J_4^2-J_5^2+2J_4J_5)&=&0,\nonumber\\
J_2(J_3^2+2J_1J_3-4J_5^2+4J_4J_5)&=&0.
\end{eqnarray}

Taking at the first $J_2=0$ one gets from (143) $J_5=0$. This case
with degenerate one-magnon dispersion is of poor physical interest
and was already studied in \cite{26}. Taking now $J_2\neq0$ and
using (32) one can subdivide the system (143) on two subsystems,
\begin{eqnarray}
(\Delta_2-\Delta_0)(\Delta_2-\Delta_1)(3\Delta_1-2\Delta_2-\Delta_0)&=&0,\nonumber\\
(\Delta_2-\Delta_0)[9(\Delta_1-\Delta_2)^2-4(\Delta_2-\Delta_0)^2+9]&=&0,\nonumber\\
(3\Delta_1-\Delta_2-2\Delta_0)[9(\Delta_1-\Delta_2)^2+4(\Delta_2-\Delta_0)^2-9]&=&0,
\end{eqnarray}
and
\begin{eqnarray}
(\Delta_1-\Delta_2)(3\Delta_r+3\Delta_1+\Delta_2-\Delta_0)&=&0,\nonumber\\
(\Delta_2-\Delta_0)(3\Delta_r+\Delta_2-3\Delta_1+5\Delta_0)&=&0,\nonumber\\
3\Delta_r(3\Delta_1+\Delta_2-\Delta_0)+9\Delta_1^2-18\Delta_1\Delta_2+\Delta_2^2+4\Delta_2\Delta_0-5\Delta_0^2&=&0,
\end{eqnarray}
where $\Delta_r\equiv J_r/(J_l-J_d)$.

We have separated (145) from (144) because the variable $J_r$
related to the term proportional to $\hat Q$ is auxiliary and
according to (23) has no affect on integrability. Nevertheless
solvability of the Yang-Baxter equation (140) for a given $H$
depends on the value of $J_r$. Therefore however it is necessary
for the system (144), (145) to be solvable a concrete value of
$J_r$ obtained from it has no affect on integrability.

The subsystem (144) has three solutions. The first one is the
solution (94) for which the subsystem (145) is also solvable. The
remaining two solutions of (144) are the following,
\begin{eqnarray}
\Delta_0=\Delta_2,\quad (\Delta_1-\Delta_2)^2&=&1,\\
\Delta_1=\Delta_2,\quad 4(\Delta_0-\Delta_2)^2&=&9.
\end{eqnarray}
A substitution of (146) into (145) shows that the latter subsystem
is solvable with respect to $\Delta_r$ only for,
\begin{equation}
\Delta_1\Delta_2=0.
\end{equation}
Together (146) and (148) result in (117) and (118).

Analogously a substitution of (147) into (145) gives,
\begin{equation}
\Delta_1\Delta_0=0.
\end{equation}
Together (147) and (149) results in (119) and (120).

Corresponding to the integrable cases $R$-matrices were already
presented in \cite{21} within the following basis in the space
${\mathbb C}^{16}$
\begin{equation}
f_{3(i-1)+j}=e_i\otimes e_j,\quad f_{9+i}=|0\rangle\otimes
e_j,\quad f_{12+i}=e_i\otimes|0\rangle,\quad
f_{16}=|0\rangle\otimes|0\rangle.
\end{equation}
Here $i,j=1,2,3$ and $e_1=|1\rangle^{+1}$, $e_2=|1\rangle^0$ and
$e_3=|1\rangle^{-1}$.

In this basis the $R$-matrix corresponding to (94) has the block
$XXZ$-type form
\begin{equation}
\check R^{(0)}(\lambda)=\left(\begin{array}{cccc}
\sinh(\lambda+\eta)I_9&0&0&0\\
0&\sinh{\eta}I_3&\sinh{\lambda}I_3&0\\
0&\sinh{\lambda}I_3&\sinh{\eta}I_3&0\\
0&0&0&\sinh{(\lambda+\eta)}
\end{array}\right).
\end{equation}
For a very special value of $\eta$ it was also presented in
\cite{27}.

In the cases (117) (for $J_r=0$) and (118) ($J_r=J_l$) the
matrices $H$ are correspondingly the normal and graded ${\mathbb
C}^4\otimes{\mathbb C}^4$ permutators ${\cal P}_4$ and $\tilde
{\cal P}_4$. (In the latter case the subspace generated by
$|0\rangle$ has the negative grading). The related $R$-matrices
have a rather simple form,
\begin{equation}
\check R^{(1,2)}(\lambda)=\eta I_{16}+\lambda H.
\end{equation}
Integrability of these models was first noted in \cite{28}. The
case (117) was intensively studied in \cite{3}.

The $R$-matrices related to (119) (for $J_r=J_l$) and (120)
($2J_r=5J_l$) also have block forms,
\begin{eqnarray}
\check R^{(3)}(\lambda)&=&\left(\begin{array}{cccc}
r(\lambda,\eta_0)&0&0&0\\
0&\sinh{\eta_0}I_3&\sinh{\lambda}I_3&0\\
0&\sinh{\lambda}I_3&\sinh{\eta_0}I_3&0\\
0&0&0&\sinh{(\lambda+\eta_0)}
\end{array}\right),\nonumber\\
\check R^{(4)}(\lambda)&=&\left(\begin{array}{cccc}
r(\lambda,\eta_0)&0&0&0\\
0&\sinh{\eta_0}I_3&\sinh{\lambda}I_3&0\\
0&\sinh{\lambda}I_3&\sinh{\eta_0}I_3&0\\
0&0&0&\sinh{(\eta_0-\lambda)}
\end{array}\right),
\end{eqnarray}
where $\sinh{\eta_0}=\sqrt{5}/2$ and,
\begin{equation}
r(\lambda,\eta_0)=\left(\begin{array}{ccccccccc}
f&0&0&0&0&0&0&0&0\\
0&f&0&0&0&0&0&0&0\\
0&0&f-g&0&g&0&-g&0&0\\
0&0&0&f&0&0&0&0&0\\
0&0&g&0&f-g&0&g&0&0\\
0&0&0&0&0&f&0&0&0\\
0&0&-g&0&g&0&f-g&0&0\\
0&0&0&0&0&0&0&f&0\\
0&0&0&0&0&0&0&0&f
\end{array}\right),
\end{equation}
($f=\sinh{(\lambda+\eta_0)}$, $g=\sinh{\lambda}$).

The matrix $r(\lambda,\eta_0)$ itself satisfies the Yang-Baxter
equation and describes the $S=1$ biquadratic spin chain. As it was
shown in \cite{29} this $R$-matrix as well as its generalization
(related to arbitrary $\eta$) are related to the Temperley-Lieb
algebra.

\section{Action of the $S_3$ group on the eigenspaces}
As it will be shown below (see Eq. (162)) the $S_3$-action (87) in
the $q$-space results in corresponding symmetry of Bethe wave
functions. The latter is useful (see Appendix A) for compact
representation of amplitudes.

First of all let us consider the case $S=0$ (which is analogous to
$S=3$). The matrix $M^{(0)}(k,{\bf q})$ possess the following
symmetry
\begin{equation}
M^{(0)}(k,\omega_j({\bf q}))=J_L^{(0)}(\omega_j)M^{(0)}(k,{\bf
q})J_R^{(0)}(\omega_j),
\end{equation}
where the matrices $J_L^{(0)}(\omega_j)$ and $J_R^{(0)}(\omega_j)$
give left and right representations of the group $S_3$:
\begin{equation}
J_L^{(0)}(\omega_i)J_L^{(0)}(\omega_j)=J_L^{(0)}(\omega_i\cdot\omega_j),\quad
J_R^{(0)}(\omega_i)J_R^{(0)}(\omega_j)=J_R^{(0)}(\omega_j\cdot\omega_i).
\end{equation}
Explicit expressions for the matrices $J_{L,R}^{(0)}(\omega_j)$
may be obtained from Eqs. (79), (156) and the following
representations for generators
\begin{eqnarray}
J_L^{(0)}(\omega_1)=\left(\begin{array}{cc}
1&{\mathbb O}_{1,5}\\
{\mathbb O}_{5,1}&\tilde I_5
\end{array}\right),&\quad& J_L^{(0)}(\omega_5)=\left(\begin{array}{cc}
\tilde I_5&{\mathbb O}_{5,1}\\
{\mathbb O}_{1,5}&1
\end{array}\right),\nonumber\\
J_R^{(0)}(\omega_1)=-\left(\begin{array}{cc}
\tilde I_2&{\mathbb O}_{2,4}\\
{\mathbb O}_{4,2}&\tilde I_4
\end{array}\right),&\quad&J_R^{(0)}(\omega_5)=-\tilde I_6.
\end{eqnarray}
Here by ${\mathbb O}_{m,n}$ we denote a $m\times n$ matrix with
all zero entries while by $\tilde I_n$ a $n\times n$ matrix with
units in the second diagonal (and all other entries equal to
zero).

Similar relations
\begin{equation}
M^{(1,2)}(k,\omega_j({\bf
q}))=J_L^{(1,2)}(\omega_j)M^{(1,2)}(k,{\bf
q})J_R^{(1,2)}(\omega_j),
\end{equation}
with
\begin{equation}
J_{L,R}^{(1)}=I_3\otimes J_{L,R}^{(0)},\quad
J_{L,R}^{(2)}=I_2\otimes J_{L,R}^{(0)},
\end{equation}
are also valid for $M^{(1,2)}(k,{\bf q})$ given by (107) and
(131).

The symmetry (158) allows to produce new solutions of the Eqs.
(105) or (129) from the known one (for Eq. (84) the result is
trivial). Indeed if
\begin{equation}
M^{(n)}(k,{\bf q})F^{(n)}(k,{\bf q})=0,
\end{equation}
for some vector $F^{(n)}(k,{\bf q})$ (${\rm dim}(F^{(1)}(k,{\bf
q}))=18$, ${\rm dim}(F^{(2)}(k,{\bf q}))=12$) then according to
(158)
\begin{equation}
M^{(n)}(k,{\bf q})J_R^{(n)}(\omega_j)F^{(n)}(k,\omega_j({\bf
q}))=0.
\end{equation}
In other words we have obtained the following action of the group
$S_3$ on the eigenspaces
\begin{equation}
\omega_j(F^{(n)})(k,{\bf
q})=J_R^{(n)}(\omega_j)F^{(n)}(k,\omega_j({\bf q})).
\end{equation}
Here $\omega_j(F^{(n)})$ is vector related to the new solution
(which in fact may coincide with the present one).

\section{Summary}

In the present paper we analyzed two- and three-magnon problems
for a rung-dimerized spin ladder. It was shown that the Bethe form
of the two-magnon solution may be obtained in a straightforward
manner from the corresponding $\rm Shr\ddot odinger$ equation.

The three-magnon problem was first analyzed in general outlook in
all sectors of total spin $S=0,1,2,3$. It was shown that at all
$S$ the reduced to the center of mass frame $\rm Shr\ddot odinger$
equation is invariant under the corresponding duality
transformation while the Fourier substitution (70) naturally
results in Bethe form of wave function.

Applicability of the Bethe Ansatze for the three-magnon problem
was analyzed separately in all sectors of total spin. It was shown
that for $S=0$ and $S=3$ the problem is always solvable and the
corresponding solution has form typical to the $XXZ$ model. The
sector $S=1$ is completely solvable in the five cases (94) and
(117)-(120). Nevertheless a special partial solution (see Appendix
A) exist for all values of the coupling constants. The sector
$S=2$ is solvable under one of the conditions (125), (137) or
(138). Explicit expressions for the solutions are presented in the
Appendix.

The result was compared with the previous consideration based on
an analysis of solvability of the Yang-Baxter equation. It was
shown that the three-magnon problem for a Hamiltonian $\hat H$ is
completely solvable within the Coordinate Bethe Ansatze if and
only if the corresponding $R$-matrix exist for some Hamiltonian in
the orbit $\hat H+\alpha\hat Q$ ($\alpha$ is real).

Finitely it is shown that the $S_3$-symmetry of the Bethe Ansatze
equations results in the action (162) of the group $S_3$ on the
space of Bethe vectors.

The author is very grateful to P.~P. Kulish and M.~I. Vyasovsky
for helpful discussions.

\appendix
\section{Partial solution in the $S=1$ sector}
An explicit form of the special partial solution of Eq. (105)
obtained by MAPLE is rather complicated. For example the
expressions for $B_j(k,{\bf q})$ at $j=1,..,6$ and $j=13,..,18$
contain 1106 terms while the expression for $B_j(k,{\bf q})$ at
$j=7,..,12$ contain 1090.

Since this solution is in general a single one it must be
$S_3$-symmetric and auto- (or anti-auto) dual. It may be readily
proved that these symmetry properties allow to obtain all
components from $B_1(k,{\bf q})$ and $B_7(k,{\bf q})$ using Eqs.
(104) and (162). Below we give representations for these two
components.

First of all $B_1(k,{\bf q})$ possess the following decomposition
\begin{equation}
B_1(k,{\bf q})=B^{(s)}_1(k,{\bf q})+B^{(a)}_1(k,{\bf q}),
\end{equation}
where the term $B^{(s)}_1(k,{\bf q})$ is symmetric under the
transposition,
\begin{equation}
k\rightarrow-k,\quad q_1\leftrightarrow q_2,
\end{equation}
while $B^{(a)}_1(k,{\bf q})$ is antisymmetric.

For $B^{(s)}_1(k,{\bf q})$ we have the following representation
\begin{eqnarray}
B^{(s)}_1(k,{\bf
q})&=&\frac{45}{2}F(\Delta_1,\Delta_2,\Delta_0)+\frac{45}{2}F(\Delta_2,\Delta_0,\Delta_1)+
\frac{9}{2}F(\Delta_0,\Delta_1,\Delta_2)\nonumber\\
&+&u_1u_2u_3\Big[W_1u_1u_2u_3+W_2\frac{u_1u_2}{z_1z_2}+
W_3\Big(\frac{u_1z_2}{z_1^2}+\frac{u_2z_1}{z_2^2}\Big)u_3\nonumber\\
&+&W_4\frac{u_3}{z_1z_2}+W_5\Big(\frac{u_1}{z_1^3}+\frac{u_2}{z_2^3}\Big)+\frac{W_6}{z_1^2z_2^2}\Big],
\end{eqnarray}
where
\begin{eqnarray}
F(\Delta,\Delta',\Delta'')&=&(\Delta'-\Delta)Z(k,{\bf
q},\Delta'')Z(k,\omega_2({\bf q}),\Delta'')Z(k,\omega_4({\bf
q}),\Delta'')\nonumber\\
&\cdot&[u_1u_2u_3+\Delta\Delta'\cos{k}+\Delta\Delta'(\Delta+\Delta')].
\end{eqnarray}
The parameters
\begin{eqnarray}
u_1&=&\cos{\Big(\frac{k}{3}+\frac{\tilde q_1}{2}\Big)},\quad
u_2=\cos{\Big(\frac{k}{3}-\frac{\tilde q_2}{2}\Big)},\nonumber\\
u_3&=&\cos{\Big(\frac{k}{3}+\frac{\tilde q_2-\tilde
q_1}{2}\Big)},\quad z_j={\rm e}^{i\tilde q_j/2},
\end{eqnarray}
have more simple form being expressed from $\tilde q_{1,2}$.

The coefficients $W_j$ for $j=1,2,3$ are the following
\begin{eqnarray}
W_1&=&27\Delta_1^3-5\Delta_2^3+8\Delta_0^3+45\Delta_1^2\Delta_2-75\Delta_1\Delta_2^2-90\Delta_2^2\Delta_0\nonumber\\
&+&60\Delta_2\Delta_0^2-18\Delta_1^2\Delta_0-12\Delta_1\Delta_0^2+60\Delta_1\Delta_2\Delta_0,\nonumber\\
W_2&=&18\Delta_1^3\Delta_2-10\Delta_1\Delta_2^3-45\Delta_1^3\Delta_0-50\Delta_1\Delta_0^3+15\Delta_2^3\Delta_0+
42\Delta_2\Delta_0^3\nonumber\\
&+&30(3\Delta_1^2-\Delta_2^2)\Delta_0^2+3\Delta_1\Delta_2\Delta_0(65\Delta_2-39\Delta_1-36\Delta_0),\nonumber\\
W_3&=&W_2+15(\Delta_1-\Delta_2)(\Delta_1-\Delta_0)(\Delta_2-\Delta_0)(4\Delta_0-3\Delta_1-\Delta_2).
\end{eqnarray}
For $j=4,5,6$ they may be obtained from (168) by the following
formulas (observed purely empirically)
\begin{equation}
W_4=\frac{\varphi(W_2)}{\Delta_1\Delta_2\Delta_0},\quad
W_5=\frac{\varphi(W_3)}{\Delta_1\Delta_2\Delta_0},\quad
W_6=\varphi(W_1),
\end{equation}
where the homomorphism $\varphi$ is defined as follows,
\begin{equation}
\varphi(\Delta_j)=\frac{\Delta_1\Delta_2\Delta_0}{\Delta_j}.
\end{equation}

For $B^{(a)}_1(k,{\bf q})$ we found the following representation
\begin{equation}
B^{(a)}_1(k,{\bf
q})=\frac{15}{2}(\Delta_1-\Delta_2)(\Delta_1-\Delta_0)(\Delta_2-\Delta_0)u_3\tilde
B^{(a)}_1(k,{\bf q}),
\end{equation}
where
\begin{eqnarray}
\tilde B^{(a)}_1(k,{\bf
q})&=&u_1u_2\Big[(3-4\Delta_1\Delta_2-6\Delta_0\Delta_2-2\Delta_0\Delta_1)\Big(\frac{u_1}{z_1^3}-\frac{u_2}{z_2^3}\Big)
\nonumber\\
&+&3i\sin{k}
+3i(1+4(\Delta_0\Delta_1+\Delta_1\Delta_2+\Delta_0\Delta_2))\frac{v_3}{z_1z_2}\nonumber\\
&-&2i(2\Delta_0+3\Delta_1+\Delta_2)\Big(\frac{u_1z_2}{z_1^2}+\frac{u_2z_1}{z_2^2}\Big)v_3\nonumber\\
&-&2i(\Delta_0+2\Delta_2)\frac{u_1v_2+u_2v_1}{z_1z_2}\Big]-6i\Delta_0\Delta_1\Delta_2\frac{u_1v_2+u_2v_1}{z^2_1z^2_2},
\end{eqnarray}
and
\begin{equation}
v_1=\sin{\Big(\frac{k}{3}+\frac{\tilde q_1}{2}\Big)},\quad
v_2=\sin{\Big(\frac{k}{3}-\frac{\tilde q_2}{2}\Big)},\quad
v_3=\sin{\Big(\frac{k}{3}+\frac{\tilde q_2-\tilde q_1}{2}\Big)}.
\end{equation}

Representation of $B_7(k,{\bf q})$ is similar to (165)
\begin{eqnarray}
B_7(k,{\bf
q})&=&45F(\Delta_0,\Delta_2,\Delta_1)+27F(\Delta_0,\Delta_1,\Delta_2)\nonumber\\
&+&u_1u_2u_3\Big[\tilde W_1u_1u_2u_3+\tilde
W_2\frac{u_1u_2}{z_1z_2}+
\tilde W_3\Big(\frac{u_1z_2}{z_1^2}+\frac{u_2z_1}{z_2^2}\Big)u_3\nonumber\\
&+&\tilde W_4\frac{u_3}{z_1z_2}+\tilde
W_5\Big(\frac{u_1}{z_1^3}+\frac{u_2}{z_2^3}\Big)+\frac{\tilde
W_6}{z_1^2z_2^2}\Big],
\end{eqnarray}
where
\begin{eqnarray}
\tilde
W_1&=&27\Delta_1^3+5\Delta_2^3-32\Delta_0^3+45\Delta_1\Delta_2(\Delta_2-\Delta_1)\nonumber\\
&+&72\Delta_1\Delta_0(\Delta_1-\Delta_0)+120\Delta_2\Delta_0(\Delta_2-\Delta_0),\nonumber\\
\tilde
W_2&=&45\Delta_1^3\Delta_0-72\Delta^3_1\Delta_2-80\Delta_2^3\Delta_1+75\Delta_2^3\Delta_0+20\Delta_0^3\Delta_1+
12\Delta_0^3\Delta_2\nonumber\\
&+&60(2\Delta_1^2\Delta_2^2-\Delta_1^2\Delta_0^2-\Delta_2^2\Delta_0^2)+
3\Delta_1\Delta_2\Delta_0(104\Delta_0-29\Delta_1-75\Delta_2),\nonumber\\
\tilde
W_3&=&W_2+90(\Delta_1-\Delta_2)^2(\Delta_1-\Delta_0)(\Delta_2-\Delta_0).
\end{eqnarray}
Again the parameters $\tilde W_{4,5,6}$ may be obtained from
$\tilde W_{1,2,3}$ according to (169) and (170).

\section{Additional $S=1$ solutions}

We shall use here the following notations
\begin{eqnarray}
m_1(k,{\bf q},\Delta)&=&Z(k,\omega_5({\bf q}),\Delta),\quad
m_2(k,{\bf q},\Delta)=Z(k,{\bf q},\Delta),\nonumber\\
m_3(k,{\bf q},\Delta)&=&=Z(k,\omega_1({\bf q}),\Delta),\quad
m_4(k,{\bf q},\Delta)=Z(k,\omega_2({\bf q}),\Delta),\nonumber\\
m_5(k,{\bf q},\Delta)&=&Z(k,\omega_3({\bf q}),\Delta),\quad
m_6(k,{\bf q},\Delta)=Z(k,\omega_4({\bf q}),\Delta),
\end{eqnarray}
(for definition of $Z(k,{\bf q},\Delta)$ and $\omega_j({\bf q})$
see (86) and (87)).

For $\Delta_0=\Delta_2=1$, $\Delta_1=0$ the space of additional to
(95), (101) solutions is generated by the vector
\begin{eqnarray}
B^{(1)}_1(k,{\bf q})&=&2m_4(k,{\bf
q},1)m_6(k,{\bf q},1)\sin{\frac{k+2q_1+4q_2}{6}}\sin{\frac{k-4q_1-2q_2}{6}},\nonumber\\
B^{(1)}_2(k,{\bf q})&=&2m_1(k,{\bf
q},1)m_6(k,{\bf q},1)\sin{\frac{k+2q_1+4q_2}{6}}\sin{\frac{k-4q_1-2q_2}{6}},\nonumber\\
B^{(1)}_3(k,{\bf q})&=&-m_1(k,{\bf q},1)m_2(k,{\bf q},1)m_6(k,{\bf
q},1),\nonumber\\
B^{(1)}_4(k,{\bf q})&=&-m_1(k,{\bf q},1)m_2(k,{\bf q},1)m_3(k,{\bf
q},1),\nonumber\\
B^{(1)}_5(k,{\bf q})&=&2m_2(k,{\bf
q},1)m_3(k,{\bf q},1)\sin{\frac{k+2q_1+4q_2}{6}}\sin{\frac{k+2q_1-2q_2}{6}},\nonumber\\
B^{(1)}_6(k,{\bf q})&=&2m_3(k,{\bf
q},1)m_5(k,{\bf q},1)\sin{\frac{k+2q_1+4q_2}{6}}\sin{\frac{k+2q_1-2q_2}{6}},\nonumber\\
B^{(2)}_1(k,{\bf
q})&=&2i\sin{(q_1+q_2)}\sin{\frac{k+2q_1+4q_2}{6}}\sin{\frac{k+2q_1-2q_2}{6}}m_6(k,{\bf
q},1),\nonumber\\
B^{(2)}_2(k,{\bf q})&=&-im_1(k,{\bf q},1)m_6(k,{\bf
q},1)\sin{(q_1+q_2)},\nonumber\\
B^{(2)}_5(k,{\bf q})&=&-im_2(k,{\bf q},1)m_3(k,{\bf
q},1)\sin{q_2},\nonumber\\
B^{(2)}_6(k,{\bf q})&=&2im_3(k,{\bf
q},1)\sin{q_2}\sin{\frac{k+2q_1+4q_2}{6}}\sin{\frac{k-4q_1-2q_2}{6}},\nonumber\\
B^{(3)}_1(k,{\bf q})&=&m_6(k,{\bf
q},1)\sin{q_2}\sin{(q_1+q_2)},\nonumber\\
B^{(3)}_6(k,{\bf q})&=&m_3(k,{\bf
q},1)\sin{q_2}\sin{(q_1+q_2)},\nonumber\\
B^{(j)}_l(k,{\bf q})&=&0,\quad (j,l)=(2,3-4),\,(3,2-5),
\end{eqnarray}
and its dual.

For $\Delta_0=\Delta_2=0$, $\Delta_1=1$ the space of additional to
(95), (101) solutions is generated by the vector
\begin{eqnarray}
B^{(1)}_1(k,{\bf q})&=&B^{(1)}_2(k,{\bf q})=m_5(k,{\bf
q},1)m_6(k,{\bf
q},1),\nonumber\\
B^{(1)}_3(k,{\bf q})&=&B^{(1)}_4(k,{\bf q})=-2m_6(k,{\bf
q},1)\sin{\frac{k+2q_1+4q_2}{6}}\sin{\frac{k-4q_1-2q_2}{6}},\nonumber\\
B^{(1)}_5(k,{\bf q})&=&B^{(1)}_6(k,{\bf q})=-2m_5(k,{\bf q},1)\sin{\frac{k+2q_1-2q_2}{6}}\sin{\frac{k-4q_1-2q_2}{6}},\nonumber\\
B^{(2)}_3(k,{\bf q})&=&im_6(k,{\bf q},1)\sin{(q_1+q_2)},\nonumber\\
B^{(2)}_4(k,{\bf
q})&=&-2i\sin{(q_1+q_2)}\sin{\frac{k+2q_1-2q_2}{6}}\sin{\frac{k-4q_1-2q_2}{6}}\nonumber\\
B^{(2)}_5(k,{\bf q})&=&-2i\sin{q_1}\sin{\frac{k+2q_1+4q_2}{6}}\sin{\frac{k-4q_1-2q_2}{6}},\nonumber\\
B^{(2)}_6(k,{\bf q})&=&im_5(k,{\bf q},1)\sin{q_1},\nonumber\\
B^{(3)}_4(k,{\bf q})&=&B^{(3)}_5(k,{\bf
q})=-\sin{q_1}\sin{(q_1+q_2)},\nonumber\\
B^{(j)}_l(k,{\bf q})&=&0,\quad (j,l)=(2,1-2),\,(3,1-3),\,(3,6),
\end{eqnarray}
and its dual.

For $\Delta_1=\Delta_2=3/2$, $\Delta_0=0$ the space of additional
to (95), (102) solutions is generated by the vector
\begin{eqnarray}
B^{(1)}_2(k,{\bf q})&=&-2m_5(k,{\bf
q},3/2)\sin{q_1}\sin{q_2},\nonumber\\
B^{(1)}_3(k,{\bf q})&=&-2m_2(k,{\bf
q},3/2)\sin{q_1}\sin{q_2},\nonumber\\
B^{(2)}_1(k,{\bf q})&=&m_4(k,{\bf q},3/2)m_5(k,{\bf
q},3/2)m_6(k,{\bf
q},3/2),\nonumber\\
B^{(2)}_2(k,{\bf q})&=&m_1(k,{\bf q},3/2)m_5(k,{\bf
q},3/2)m_6(k,{\bf
q},3/2),\nonumber\\
B^{(2)}_3(k,{\bf q})&=&m_1(k,{\bf q},3/2)m_2(k,{\bf
q},3/2)m_6(k,{\bf
q},3/2),\nonumber\\
B^{(2)}_4(k,{\bf q})&=&m_1(k,{\bf q},3/2)m_2(k,{\bf
q},3/2)m_3(k,{\bf
q},3/2),\nonumber\\
B^{(2)}_5(k,{\bf q})&=&-B^{(3)}_5(k,{\bf q})=m_2(k,{\bf
q},3/2)m_3(k,{\bf q},3/2)m_4(k,{\bf
q},3/2),\nonumber\\
B^{(2)}_6(k,{\bf q})&=&-B^{(3)}_6(k,{\bf q})=m_3(k,{\bf
q},3/2)m_4(k,{\bf q},3/2)m_5(k,{\bf
q},3/2),\nonumber\\
B^{(3)}_1(k,{\bf q})&=&-m_4(k,{\bf q},3/2)m_5(k,{\bf
q},3/2)\Big[\cos{\Big(\frac{k-q_1-2q_2}{3}\Big)}-\cos{q_1}-\frac{{\rm
e}^{iq_1}}{2}\Big],\nonumber\\
B^{(3)}_2(k,{\bf q})&=&-m_1(k,{\bf q},3/2)m_5(k,{\bf
q},3/2)\Big[\cos{\Big(\frac{k-q_1-2q_2}{3}\Big)}-\cos{q_1}-\frac{{\rm
e}^{iq_1}}{2}\Big],\nonumber\\
B^{(3)}_3(k,{\bf q})&=&-m_2(k,{\bf q},3/2)m_6(k,{\bf
q},3/2)\Big[\cos{\Big(\frac{k+2q_1+q_2}{3}\Big)}-\cos{q_2}-\frac{{\rm
e}^{-iq_2}}{2}\Big],\nonumber\\
B^{(3)}_4(k,{\bf q})&=&-m_2(k,{\bf q},3/2)m_3(k,{\bf
q},3/2)\Big[\cos{\Big(\frac{k+2q_1+q_2}{3}\Big)}-\cos{q_2}-\frac{{\rm
e}^{-iq_2}}{2}\Big],\nonumber\\
B_j(k,{\bf q})&=&0,\quad (j,l)=(1,1),\,(1,4-6),
\end{eqnarray}
and its dual.

For $\Delta_0=-3/2$, $\Delta_1=\Delta_2=0$ the space of additional
to (95), (102) solutions is generated by the vector
\begin{eqnarray}
B^{(1)}_1(k,{\bf q})&=&B^{(1)}_6(k,{\bf q})=-im_5(k,{\bf
q},-3/2)m_6(k,{\bf
q},-3/2)\sin{q_2},\nonumber\\
B^{(1)}_4(k,{\bf q})&=&B^{(1)}_5(k,{\bf q})=im_1(k,{\bf
q},-3/2)\sin{q_1}\nonumber\\
&\cdot&\Big(\cos{\frac{k+q_2-q_1}{3}+\cos{(q_1+q_2)}}+\frac{{\rm
e}^{-i(q_1+q_2)}}{2}\Big),\nonumber\\
B^{(2)}_l(k,{\bf q})&=&m_1(k,{\bf q},-3/2)m_5(k,{\bf
q},-3/2)m_6(k,{\bf
q},-3/2),\quad l=1,2,3,4,5,6,\nonumber\\
B^{(3)}_3(k,{\bf q})&=&B^{(3)}_4(k,{\bf q})=im_1(k,{\bf
q},-3/2)m_6(k,{\bf
q},-3/2)\sin{(q_1+q_2)},\nonumber\\
B^{(3)}_5(k,{\bf q})&=&B^{(3)}_6(k,{\bf q})=im_5(k,{\bf
q},-3/2)\sin{q_1}\nonumber\\
&\cdot&\Big(\cos{\frac{k+2q_1+q_2}{3}}+\cos{q_2}+\frac{{\rm
e}^{iq_2}}{2}\Big),\nonumber\\
B^{(j)}_l(k,{\bf q})&=&0,\quad (j,l)=(1,2-3),\,(3,1-2),
\end{eqnarray}
and its dual.

\section{$S=2$ solutions}

For $\Delta_1=\pm1$, $\Delta_2=0$ the space of solutions is
spanned on
\begin{eqnarray}
C^{(1)}_1(k,{\bf q})&=&C^{(1)}_6(k,{\bf q})=-im_5(k,{\bf
q})m_6(k,{\bf
q},\pm1)\sin{q_2},\nonumber\\
C^{(1)}_2(k,{\bf q})&=&C^{(1)}_3(k,{\bf q})=0,\nonumber\\
C^{(1)}_4(k,{\bf q})&=&C^{(1)}_5(k,{\bf q})=im_1(k,{\bf
q},\pm1)m_2(k,{\bf
q},\pm1)\sin{q_1},\nonumber\\
C^{(2)}_1(k,{\bf q})&=&C^{(2)}_2(k,{\bf q})=\pm m_1(k,{\bf
q},\pm1)m_5(k,{\bf
q},\pm1)m_6(k,{\bf q},\pm1),\nonumber\\
C^{(2)}_3(k,{\bf q})&=&C^{(2)}_4(k,{\bf q})=\pm m_1(k,{\bf
q},\pm1)m_2(k,{\bf
q},\pm1)m_6(k,{\bf q},\pm1),\nonumber\\
C^{(2)}_5(k,{\bf q})&=&C^{(2)}_6(k,{\bf q})=\pm m_2(k,{\bf
q},\pm1)m_5(k,{\bf q},\pm1)\Big({\rm
e}^{i(k+2q_1-2q_2)/6}\nonumber\\
&\mp&{\rm e}^{-i(k+2q_1-2q_2)/6}\Big),
\end{eqnarray}
and its dual.

For $\Delta_1=0$, $\Delta_2=\pm1$ the space of solutions is
spanned on
\begin{eqnarray}
C^{(1)}_1(k,{\bf q})&=&C^{(1)}_6(k,{\bf q})=0,\nonumber\\
C^{(1)}_2(k,{\bf q})&=&im_5(k,{\bf q},\pm1)m_6(k,{\bf
q},\pm1)\sin{q_2},\nonumber\\
C^{(1)}_3(k,{\bf q})&=&im_2(k,{\bf q},\pm1)m_6(k,{\bf
q},\pm1)\sin{q_2},\nonumber\\
C^{(1)}_4(k,{\bf q})&=&im_1(k,{\bf q},\pm1)m_6(k,{\bf
q},\pm1)\sin{(q_1+q_2)},\nonumber\\
C^{(1)}_5(k,{\bf q})&=&im_4(k,{\bf q},\pm1)m_6(k,{\bf
q},\pm1)\sin{(q_1+q_2)},\nonumber\\
C^{(2)}_1(k,{\bf q})&=&C^{(2)}_6(k,{\bf q},\pm1)=\mp m_4(k,{\bf
q},\pm1)m_5(k,{\bf
q},\pm1)m_6(k,{\bf q},\pm1),\nonumber\\
C^{(2)}_2(k,{\bf q})&=&\mp m_1(k,{\bf q},\pm1)m_5(k,{\bf
q},\pm1)m_6(k,{\bf q},\pm1),\nonumber\\
C^{(2)}_3(k,{\bf q})&=&\mp m_6^2(k,{\bf q},\pm1)\Big({\rm
e}^{i(k+2q_1+4q_2)/6}\nonumber\\
&\mp&{\rm e}^{-i(k+2q_1+4q_2)/6}\Big),\nonumber\\
C^{(2)}_4(k,{\bf q})&=&\mp m_3(k,{\bf q},\pm1)m_6(k,{\bf
q},\pm1)\Big({\rm
e}^{i(k+2q_1+4q_2)/6}\nonumber\\
&\mp&{\rm e}^{-i(k+2q_1+4q_2)/6}\Big),\nonumber\\
C^{(2)}_5(k,{\bf q})&=&\mp m_2(k,{\bf q},\pm1)m_4(k,{\bf
q},\pm1)m_6(k,{\bf q},\pm1),
\end{eqnarray}
and its dual.


\begin{thebibliography}{29}
\bibitem{1} Dagotto E 1999 {\it Rep. Progr. Phys.} {\bf62} 1525
\bibitem{2} Schmidt K~P, Uhrig G~S 2005 {\it Mod. Phys. Lett.} B
{\bf19} 1179
\bibitem{3} Batchelor M~T, Guan X-W, Oelkers N, Tsuboi Z 2007
{\it Adv. Phys.} {\bf 56} 465
\bibitem{4} Barnes T, Dagotto E, Riera J, Swanson E~S 1993 {\it Phys.
Rev.} B {\bf 47} 3196
\bibitem{5} Brehmer S, Mikeska H-J, and $\rm M\ddot uller$ M,
Nagaosa N and Uchida S 1999 {\it Phys. Rev.} B {\bf 60} 329-334
\bibitem{6} Kolezhuk A~K and Mikeska H-J 1998 {\it Int. J. Mod.
Phys.} B {\bf 12} 2325
\bibitem{7} Bibikov P~N 2005 {\it Phys. Rev.} B {\bf72} 012416
\bibitem{8} Bibikov P~N 2006 {\it Phys. Rev.} B {\bf73} 132402
\bibitem{9} Bibikov P~N and Vyazovsky M~I 2007 {\it Phys. Rev.} B {\bf75} 094420
\bibitem{10} Bibikov P~N 2007 {\it Phys. Rev.} B {\bf76} 174431
\bibitem{11} Essler F~H~L 2000 {\it Phys. Rev.} B {\bf 62} 3264
\bibitem{12} Kirillov A~N, Smirnov F~A 1998 {\it Int. Journ. Mod.
Phys.} A {\bf 3} 731
\bibitem{13} Gaudin M 1983 {\it La Fonction D'onde de Bethe} (Paris: Masson)
\bibitem{14} Babbit D, Gutkin E 1990 {\it Lett. Math. Phys.}
{\bf20} 91
\bibitem{15} Faddeev L~D 1998 {\it How algebraic Bethe Ansatz works for integrable
models}, Quantum symmetries/Symmetries quantique, {\it Proceedings
of the Les Houches summer school} Session LXIV, eds. A. Connes, K.
Gawedzki and J. Zinn-Justin  North-Holland
\bibitem{16} Korepin V~E, Izergin A~G, Bogoliubov N~M 1993
{\it Quantum inverse scattering method and correlation functions}
(Cambridge: Univ. Press)
\bibitem{17} Kulish P~P, Sklyanin E~K 1982 {\it Quantum spectral transform method. Recent
developments.} {\it Proc. Symp. on Integrable Quantum Fields} {\it
Lecture Notes in Physics 151} Eds. J.~Hietarinta and C.~Montonen
(New York: Springer)
\bibitem{18} Hess C 2008 {\it Eur. Phys. J. Special Topics} {\bf 151}
73-83
\bibitem{19} ${\rm M\ddot utter}$K-H, Schmitt A 1995 {\it J. Phys. A: Math. Gen.} {\bf28} 2265
\bibitem{20} Bibikov P~N 2003 {\it Phys. Lett.} A {\bf314} 209
\bibitem{21} Bibikov P~N 2003 {\it Zap. Nauchn. Semin. POMI} {\bf291} 24
\bibitem{22} Bibikov P~N 2007 {\it J. Phys.} A {\bf 40} 4683
\bibitem{23} Gritsev V, Baeriswyl D 2003 {\it J. Phys. A: Math.
Gen.} {\bf36} 12129-12142
\bibitem{24} An idea to use a three-magnon problem as an
alternative integrability test was prompted to the author by P.~P.
Kulish.
\bibitem{25} Sachdev S, Bhatt R~N 1990 {\it Phys. Rev.} B {\bf41} 9323
\bibitem{26} Legeza $\rm \ddot O$, $\rm F\acute ath$ G, $\rm S\acute
olyom$ J 1997 {\it Phys. Rev.} B {\bf55} 291-298
\bibitem{27} Albeverio S, Fei S-M and Wang Y 1999 {\it Europhys. Lett.}, {\bf
47} 364-370
\bibitem{28} Wang Y 1999
{\it Phys. Rev. B} {\bf 60} 9236-9239
\bibitem{29} Kulish P~P 2003
{\it J. Phys. A: Math. Gen.} {\bf36} L489-L493
\end{thebibliography}
\end{document}